\newif\ifjcp
\jcptrue  

\ifjcp
    \documentclass[twocolumn,aip,jcp,reprint,floatfix]{revtex4-1}
\else
    \documentclass[journal=jacsat,manuscript=article]{achemso}
    \setkeys{acs}{articletitle = true}
    \newcommand{\onlinecite}[1]{\hspace{-1 ex} \nocite{#1}\citenum{#1}} 
    \SectionNumbersOn
\fi

\usepackage{float}
\usepackage{amsmath}
\usepackage{physics}
\usepackage[dvipsnames,table]{xcolor}
\usepackage{graphicx}
\usepackage{subfig}

\newcommand{\vtwo}[1]{{#1}}

\newcommand{\setinfo}{%
    \title{Constructing `full-frequency' spectra via moment constraints for coupled cluster Green's functions}
    \author{Oliver J. Backhouse}%
    \affiliation{Department of Physics, King's College London, Strand, London WC2R 2LS, U.K.}%
    \author{George H. Booth}%
    \email{george.booth@kcl.ac.uk}
    \affiliation{Department of Physics, King's College London, Strand, London WC2R 2LS, U.K.}%
}

\ifjcp
    \renewenvironment{figure}{%
        \begin{figure*}%
    }{%
        \end{figure*}%
        \ignorespacesafterend%
    }
\else
\fi

\makeatletter
\def\@firstoftwo@second#1#2{%
  \def\temp##1.##2\@nil{##2}%
   \temp#1\@nil}
\newcommand\sref[1]{\expandafter\@setref\csname r@#1\endcsname\@firstoftwo@second{#1}}
\makeatother

\ifjcp\else
    \setinfo
\fi

\begin{document}

\ifjcp
    \setinfo
\fi

\begin{abstract}
We propose an approach to build `full-frequency' quasiparticle spectra from conservation of a set of static expectation values. These expectation values define the moments of the spectral distribution, resulting in an efficient and systematically improvable expansion. By computing these initial moment constraints at the coupled-cluster level, we demonstrate convergence in both correlated state-specific and full spectral quantities, while requiring a fraction of the effort of traditional Green's function approaches. Tested across the GW100 benchmark set for charged excitation spectra, we can converge frontier excitations to within the inherent accuracy of the CCSD approximation, whilst obtaining a simultaneous representation of the entire excitation spectrum at all energy scales.
\end{abstract}

\ifjcp
    \maketitle
\fi

\newcommand{\melpsi}[1]{\mel{\mathbf{\Psi}}{#1}{\mathbf{\Psi}}}
\newcommand{\melphi}[1]{\mel{\mathbf{\Phi}}{#1}{\mathbf{\Phi}}}
\newcommand{\cre}{a^\dagger}
\newcommand{\des}{a}
\newcommand{\cres}{\bar{a}^\dagger}
\newcommand{\dess}{\bar{a}}


\section{Introduction}

The single-particle spectrum of a quantum many-body system characterizes the position and weight of all the occupied and unoccupied energy levels present,
and its central nature in rationalizing electronic structure means that it is also referred to as the `fundamental' spectrum.
This ensures its prominence in numerical methods in electronic structure, and accurately and tractably describing the effect of electron correlation on this quantity is an open and challenging research area \cite{%
    Hedin1965,
    VanSetten2013,
    VonNiessen1984,
    VanSchilfgaarde2005,
    Schirmer1983,
    Trofimov1995,
    Trofimov1999,
    Onida2002,
    Iskakov2019,
    Metzner1989,
    Zhang1993,
    Georges1996,
    VanNeck2001,
    Backhouse2020b,
    Backhouse2021
}.
This quantity is directly probed experimentally via photoelectron spectroscopy,
as well as other techniques such as scanning tunnelling microscopy \cite{Doyen1993, Lin2010, Golze2019, Himpsel1983}.
These quantities can be modelled theoretically according to the single particle Green's function,
$G$,
expressed as a matrix-valued function of a continuous variable in the real-frequency domain.
For an $N$-electron system described by a shifted Hamiltonian $H_N = H - E_0$,
the Green's function can also be expressed in the time domain as
\begin{align}
    G_{pq}(\tau)
    &=
    \melpsi{
        \cre_{q}
        e^{-2 \pi i H_N \tau}
        \des_{p}
    }
    \nonumber
    \\
    &+
    \melpsi{
        \des_{p}
        e^{2 \pi i H_N \tau}
        \cre_{q}
    }
    ,
    \label{eq:greens_function_time}
\end{align}
where the parameter $\tau = t - t^\prime$,
indicating the dependence only on the difference in time for evolution under a time-independent Hamiltonian.
The first and second term correspond to the hole (direct photoelectron) and particle (inverse photoelectron) Green's functions,
respectively.
The Fourier transform of Eq.~\ref{eq:greens_function_time} yields the corresponding expression in the frequency domain
\begin{align}
    G_{pq}(\omega)
    &=
    \melpsi{
        \cre_{q}
        [\omega + H_N + i\eta]^{-1}
        \des_{p}
    }
    \nonumber
    \\
    &+
    \melpsi{
        \des_{p}
        [\omega - H_N + i\eta]^{-1}
        \cre_{q}
    }
    ,
    \label{eq:greens_function_freq}
\end{align}
where $\eta$ is a small positive broadening factor required to formally regularize this Fourier transform, and defines this (retarded) Green's function.
The fundamental spectrum defining the density of states of the system,
as well as the fundamental (quasiparticle) gap,
is then characterized as
\begin{equation}
    \label{eq:spectrum}
    A(\omega) = - \frac{1}{\pi} \Tr [ \Im G(\omega)],
\end{equation}
which tends to a series of delta functions at all ionization potentials and electron affinities of the correlated $N-$body interacting system as $\eta \rightarrow 0^+$.

At the mean-field level of theory,
the ground-state $\ket{\mathbf{\Psi}}$ is defined by a single (typically Hartree--Fock) reference state determinant $\ket{\mathbf{\Phi}}$,
and in the limit of vanishing $\eta$,
the spectrum of $G_{pq}(\omega)$ consists of a series of equally-weighted delta functions at the orbital energies of the single-particle Hamiltonian.
In this work, we aim to include the effects of correlation on this spectrum,
which introduces additional peaks in the spectrum,
modifications and rearrangement of spectral intensity between the peaks,
and shifting of the position of these energy levels,
with important modifications for the described ionization potentials (IPs),
electron affinities (EAs) and fundamental gaps of the system \cite{Fetter2003}.
Specifically,
we aim to describe these correlation-driven changes to the spectrum via the coupled-cluster (CC) level of theory\cite{Cizek1966, Cizek1969, Cizek1971, Paldus1972, Purvis1982, Hirata2004, Bartlett2007}.
This theory is well-known as the gold standard of quantum chemistry,
and has recently seen a revival in its use for describing spectral functions according to the CC Green's function (CCGF) formalism \cite{Nooijen1992, Nooijen1993, Nooijen1995a, BhaskaranNair2016, Shee2019, Peng2019, Peng2021a, Peng2021b},
which is closely connected to the equation of motion (EOM-CC) formalism \cite{Stanton1993, Nooijen1995b, Lange2018} and includes recent applications to the description of bandstructures in the solid state and combination with other embedded numerical methods \cite{Gruneis2011, Voloshina2011, Walz2012, Booth2013, Berkelbach2017, Furukawa2018, Shee2019, Zhu2019, Zhang2019, Zhu2021, Shee2022}.

In this work,
we describe a numerically simple and low-scaling approach to obtain a systematically improvable approximation to the coupled-cluster single-particle Green's function across all energy scales, and
which can be simply adapted from any equation-of-motion coupled-cluster code.
This allows for the full CC Green's function including off-diagonal elements to be probed for all frequencies,
without requiring {\em a priori} frequency grid definitions on which the function is resolved.
Furthermore,
this Green's function and self-energy can be directly obtained as a series of specific energies and spectral weights of all poles in the $\eta \rightarrow 0^+$ limit,
in a fashion similar to recent reformulations of `frequency-free' GW, GF2 and other correlated Greens function methods \cite{Hirata2015, Bintrim2021, Bintrim2022, Backhouse2021, Backhouse2020a, Backhouse2020b, Loos2020}.
Furthermore, in contrast to some other approaches,
this CC Green's function is not solved for one frequency at a time,
nor resolved as a state-specific expansion of successive IPs and EAs.

Instead, a series of moments of the IP and EA spectral distributions are recursively computed at the CC level,
which can be formulated as simple expectation values.
From these, a Green's function can be algebraically constructed which ensures that those occupied and unoccupied spectral moments are exactly matched. This also contrasts conceptually with common Lanczos-based approaches which formally conserve a series of moments of the \emph{self-energy}, rather than the Green's function as done here.
We show that this approach can converge a good approximation to the full frequency dynamics of the coupled-cluster Green's function whilst saving many orders of magnitude in the number of matrix-vector multiplications of the CC similarity transformed Hamiltonian,
compared to traditional approaches based on solving at individual frequencies \cite{Nooijen1993, BhaskaranNair2016, Zhu2019, Peng2019, Peng2021a, Peng2021b}.
Furthermore,
this approach is fundamentally adapted for the non-Hermitian nature of the CC Hamiltonian,
ensuring that both diagonal and off-diagonal elements of the Green's function are treated faithfully.

In section~\ref{sec:CCGF} we recapitulate the formal perspectives of coupled-cluster Green's function (CCGF) theory, describing our approach based on a truncated Green's function moment expansion in Sec.~\ref{sec:MomCCGF}.
Finally, we numerically compare the convergence of our moment expansion to the traditional `correction vector' approach to CCGF,
before analysing the convergence of CCGF for a common test set of molecular systems for spectral properties (GW100 test set \cite{VanSetten2015}) in Sec.~\ref{sec:results}.

\section{Coupled-cluster Green's functions} \label{sec:CCGF}

In this section, we review the formulation of coupled-cluster Green's function theory,
with more details available in Ref.~\onlinecite{Peng2021b}.
In coupled cluster,
the ground state is parameterised according to an explicitly size-extensive exponential ansatz \cite{Bartlett2007}, as
\begin{align}
    \label{eq:exponential_ansatz}
    \ket{\mathbf{\Psi}_\mathrm{R}}
    &=
    e^T \ket{\mathbf{\Phi}}
    ,
\end{align}
where the cluster operator $T$ can be expanded in terms of particle-hole excitations (where occupied spin-orbitals are denoted $i,j,k,\dots$ and virtual spin-orbitals as $a,b,c,\dots$) from a reference state up to a given order
\begin{align}
    T
    &=
    T_1
    +
    T_2
    +
    ...
    \nonumber
    \\
    &=
    \sum_{ia}
    t_{i}^{a} \cre_{a} \des_{i}
    +
    \sum_{i<j, a<b}
    t_{ij}^{ab} \cre_{a} \cre_{b} \des_{j} \des_{i}
    +
    ...
    \label{eq:cc_t_operator}
\end{align}
Since $e^T$ is non-unitary,
the bra corresponding to $\ket{\mathbf{\Psi_\mathrm{R}}} = e^T \ket{\mathbf{\Phi}}$ cannot be expressed simply as its adjoint,
and instead one must introduce a pair of biorthogonal wavefunctions required for general expectation values \cite{Stanton1993}, with
\begin{align}
    \label{eq:biorthogonal_system}
    \braket{\mathbf{\Psi}_\mathrm{L}}{\mathbf{\Psi}_\mathrm{R}}
    &=
    1
    .
\end{align}
Typically one expands $\bra{\mathbf{\Psi_\mathrm{L}}}$ in a linear de-excitation operator $\Lambda$
\begin{align}
    \label{eq:exponential_ansatz_bra}
    \bra{\mathbf{\Psi_\mathrm{L}}}
    &=
    \bra{\mathbf{\Phi}} (1 + \Lambda) e^{-T}
    ,
\end{align}
where the $\Lambda$ de-excitation operator can be written in a similar truncated fashion to Eq.~\ref{eq:cc_t_operator}, as
\begin{align}
    \Lambda
    &=
    \Lambda_1
    +
    \Lambda_2
    +
    ...
    \nonumber
    \\
    &=
    \sum_{ia}
    l_{a}^{i} \cre_{i} \des_{a}
    +
    \sum_{i<j, a<b}
    l_{ab}^{ij} \cre_{i} \cre_{j} \des_{b} \des_{a}
    +
    ...
    \label{eq:cc_l_operator}
\end{align}
After insertion of the biorthogonal ground state wavefunctions (Eqs.~\ref{eq:exponential_ansatz} and \ref{eq:exponential_ansatz_bra}) into the expression for the single particle Green's function (Eq.~\ref{eq:greens_function_freq}),
one arrives at the expression for the (retarded) coupled cluster Green's function
\begin{align}
    G_{pq}(\omega)
    &=
    \melphi{
        (1 + \Lambda)
        e^{-T}
        \cre_{q}
        [\omega + H_N + i\eta]^{-1}
        \des_{p}
        e^T
    }
    \nonumber
    \\
    &+
    \melphi{
        (1 + \Lambda)
        e^{-T}
        \des_{p}
        [\omega - H_N + i\eta]^{-1}
        \cre_{q}
        e^T
    }
    .
    \label{eq:greens_function_cc_expanded}
\end{align}
This can be reformulated in terms of a similarity-transformed normal-ordered Hamiltonian,
$\bar{H}_N = e^{-T} H e^{T} - E_{CC}$, where $E_{CC}$ is the ground-state CC energy.
This operator is non-Hermitian,
ensuring that the resulting Green's function is also non-Hermitian,
even in the limit $\eta \rightarrow 0$.
By noting that $e^{T} e^{-T} = I$ and also introducing the notation for similarity-transformed creation and annihilation operators
$\dess_{p} = e^{-T} \des_{p} e^{T}$
and $\cres_{p} = e^{-T} \cre_{p} e^{T}$,
Eq.~\ref{eq:greens_function_cc_expanded} can be rewritten as
\begin{align}
   G_{pq}(\omega)
    &=
    \melphi{
        (1 + \Lambda)
        \cres_{q}
        [\omega + \bar{H}_N + i\eta]^{-1}
        \dess_{p}
    }
    \nonumber
    \\
    &+
    \melphi{
        (1 + \Lambda)
        \dess_{p}
        [\omega - \bar{H}_N + i\eta]^{-1}
        \cres_{q}
    }
    ,
    \label{eq:greens_function_cc}
\end{align}
A pair of explicitly frequency-dependent many-body operators are introduced,
\begin{align}
    X_{p}(\omega)
    &=
    X_{p, 1}(\omega)
    +
    X_{p, 2}(\omega)
    +
    ...
    \nonumber
    \\
    &=
    \sum_{i}
    [x^{i} (\omega)]_{p} \des_{i}
    +
    \sum_{i<j, a}
    [x^{ij}_{a} (\omega)]_{p} \cre_{a} \des_{j} \des_{i}
    ,
    \label{eq:cc_x_operator}
    \\
    Y_{q}(\omega)
    &=
    Y_{q, 1}(\omega)
    +
    Y_{q, 2}(\omega)
    +
    ...
    \nonumber
    \\
    &=
    \sum_{a}
    [y^{a} (\omega)]_{q} \cre_{a}
    +
    \sum_{i, a<b}
    [y^{ab}_{i} (\omega)]_{q} \cre_{a} \cre_{b} \des_{i}
    ,
    \label{eq:cc_y_operator}
\end{align}
whose parameters can be optimized at each frequency point to solve the system of linear equations,
\begin{align}
    P_X (\omega + \bar{H}_N + i\eta) X_{p} \ket{\mathbf{\Phi}}
    &=
    P_X \dess_{p} \ket{\mathbf{\Phi}}
    ,
    \label{eq:cc_x_relation}
    \\
    P_Y (\omega - \bar{H}_N + i\eta) Y_{q} \ket{\mathbf{\Phi}}
    &=
    P_Y \cres_{q} \ket{\mathbf{\Phi}}
    .
    \label{eq:cc_y_relation}
\end{align}
In these expressions, $P_X$ and $P_Y$ are projection operators onto the appropriate space of electron removal/addition states to ensure well-determined equations, i.e.
\begin{align}
    P_X
    &=
    P_{X,1} + P_{X,2} + \dots
    \nonumber
    \\
    &=
    \sum_{i}
    \des_{i} \ket{\mathbf{\Phi}} \bra{\mathbf{\Phi}} \cre_{i}
    +
    \sum_{i<j, a}
    \cre_{a} \des_{j} \des_{i}
    \ket{\mathbf{\Phi}}
    \bra{\mathbf{\Phi}}
    \cre_{i} \cre_{j} \des_{a} + \dots
    ,
    \label{eq:ccsd_ip_projection}
    \\
    P_Y
    &=
    P_{Y,1} + P_{Y,2} + \dots
    \nonumber
    \\
    &=
    \sum_{a}
    \cre_{a} \ket{\mathbf{\Phi}} \bra{\mathbf{\Phi}} \des_{a}
    +
    \sum_{i, a<b}
    \cre_{a} \cre_{b} \des_{i}
    \ket{\mathbf{\Phi}}
    \bra{\mathbf{\Phi}}
    \cre_{i} \des_{b} \des_{a} + \dots
    .
    \label{eq:ccsd_ea_projection}
\end{align}
Once these equations are satisfied for a given frequency,
one may evaluate the coupled cluster Green's function as
\begin{align}
    G_{pq}(\omega)
    &=
    \melphi{
        (1 + \Lambda)
        \cres_{q}
        X_{p} (\omega)
    }
    \nonumber
    \\
    &+
    \melphi{
        (1 + \Lambda)
        \dess_{p}
        Y_{q} (\omega)
    }
    .
    \label{eq:greens_function_cc_xy}
\end{align}

The most common coupled cluster singles and doubles (CCSD) method consists of a truncation of $T$ to first- and second-order contributions only.
Equivalent truncations are then applied to the $\Lambda$ operator,
while the $X$, $Y$, $P_X$ and $P_Y$ operators are truncated to span $1h/2h1p$ and $1p/1h2p$ spaces respectively,
as shown in Eqs.~\ref{eq:cc_x_operator} and \ref{eq:cc_y_operator}.
The choice to truncate the $X$ and $Y$ operators to these spaces ensures that the fluctuation space describing the singly-charged excitation manifold is generally not complete, but instead designed to provide a consistent level of description of both the ground and excited states. However, this truncation results in some expectation values from the Green's function not matching their analogous properties from ground-state coupled cluster theory.
In particular,
the Galitskii-Migdal expression of the energy from the Green's function does not match the ground-state energy of the system at the CCSD level,
which requires dynamical fluctuations into the $3h2p$ space \cite{Nooijen1992, Nooijen1993}. This aspect will be considered further in Sec.~\ref{sec:results}.

By employing iterative linear equation solvers for Eqns.~\ref{eq:cc_x_relation} and \ref{eq:cc_y_relation},
the dominant cost in computing the Green's function is reduced to a series of matrix-vector operations between $\bar{H}_N$ and a trial vector of $X_p(\omega)$ for each frequency value of interest, where explicit expressions for the initial vectors can be found in Ref.~\onlinecite{Shee2019}, with expressions for the action of ${\hat H}_N$ found in Ref.~\onlinecite{Nooijen1995a}.
These matrix-vector products represent the core routines in any IP-EOM-CC and EA-EOM-CC implementations \cite{Stanton1993, Nooijen1995a, Nooijen1995b}.
This results in a iterative scaling with system size for CCSD-GF of $\mathcal{O}[N_\omega N_\mathrm{orb} N_\mathrm{occ}^{4} N_\mathrm{vir}]$ for the particle removal (IP) states of Eq.~\ref{eq:cc_x_relation},
and $\mathcal{O}[N_\omega N_\mathrm{orb} N_\mathrm{occ} N_\mathrm{vir}^{4}]$ for the particle addition (EA) states of Eq.~\ref{eq:cc_y_relation},
where $N_{\omega}$ is the number of frequency points,
and assuming that the full Green's function is sought.
The final step corresponding to Eq.~\ref{eq:greens_function_cc_xy} represents a lower,
non-iterative $\mathcal{O}[N_{\omega} N^5]$ scaling.

To mitigate the overall cost, a number of numerical approximations to full CCGF approaches have been developed,
as well as efficient parallel implementations.
One such implementation employs a model-order reduction technique to project the linear problem onto a more compact subspace over a chosen set of frequency intervals,
resulting in a simplified iterative solution for each frequency \cite{Peng2019}.
This method is implemented in the highly parallelized \texttt{GFCCLib} package\cite{Peng2021a}.
Shee and Zgid propose an alternative method in which the similarity-transformed Hamiltonian is recursively projected into a tridiagonal subspace by means of a biorthogonal Lanczos approach\cite{Shee2019}.
From this, the Green's function can be directly constructed and circumvents the need to explicitly solve the linear equations at each frequency.
We believe that this is equivalent to a moment expansion of the effective self-energy of the system \cite{Rusakov2014, Backhouse2020a, Sriluckshmy2021}.

Due to difficulties with the biorthogonal nature of the theory, Shee and Zgid determined the off-diagonal elements of the Green's function via computing contributions from the sum of all pairs of single-particle perturbation operators.
Formally this results in an $\mathcal{O}[N^7]$ scaling method if all elements of the full system Greens function matrix are required. However,
the authors were primarily applying the approach to compute Green's functions only over a subspace independent of system size, resulting in a return to $\mathcal{O}[N^5]$ scaling in common with other approaches (and the current work) in this case.
The significant advantage of \vtwo{these approaches} however is that it removes the dependency of the scaling with the number of frequency points,
which substantially reduces the prefactor in the Green's function construction.
We aim to retain this feature in our approach of the next section.
Finally, in the context of \textit{ab initio} solids,
Kosugi and Matsushita proposed an interpolation scheme for Green's functions which they apply to CCGF at the level of singles and doubles, to avoid the steep cost with respect to $k$-point sampling.
The scheme effectively reduces the computational load by allowing spectra on fine $k$-point meshes to be interpolated from a Fourier transform of spectra on coarser $k$-point meshes \cite{Kosugi2019}.


\section{Moment-conserved coupled-cluster Green's functions} \label{sec:MomCCGF}
\subsection{Overview}
The approach we present in this work bears similarities to recent works to formulate low-cost approaches to CCGF of Shee \textit{et al.}\cite{Shee2019} and Peng \textit{et al.}\cite{Peng2019} described above,
and inspired by aspects of the `auxiliary' GF2 approach \cite{Backhouse2020a, Backhouse2020b, Backhouse2021} and energy-weighted density matrix embedding theory\cite{Fertitta2019, Sriluckshmy2021}.
In order to remove the dependence of the CCGF with respect to an explicit frequency grid and reduce the cost,
we again solve for the coupled-cluster Green's function at all frequencies in a smaller and systematically improvable subspace compared to the full space that $\bar{H}_N$ is represented in.
This subspace is constructed independently for the hole (IP) and particle (EA) parts of the Green's functions,
and then combined to a single subspace.
A key difference to other approaches is that the coupled-cluster Hamiltonian and CCGF equations are {\em not} formally projected into an explicitly constructed subspace,
as would be common to e.g. Lanczos methods.
Instead,
a `fictitious' subspace Hamiltonian is recursively constructed,
formulated to guarantee that the particle and hole spectral moments up to a given order of the subspace Green's function exactly match the ones expected from full CCGF theory.
This therefore avoids any direct projection of $\bar{H}_N$ in the construction of the subspace Hamiltonian,
which only enters via the definition of the initial GF moments of the particle and hole spectral distributions, provided as the constraints in the subspace construction.
As the number of moment constraints increases, 
so does the accuracy of the resulting CCGF approximation, and the size of the constructed subspace.

This perspective of fictitious Hamiltonian subspace construction contrasts with the traditional Krylov subspace approach which (from a Green's function perspective) is based around conservation of spectral moments of an effective self-energy, rather than the Green's function moments here.
It has been found by the authors in Ref.~\onlinecite{Backhouse2020a}
(albeit for a different level of theory) that defining a subspace Green's function based on this conservation of the full system Green's function moments is a significantly more efficient and rapidly convergent approach than the equivalent effective self-energy truncations,
which do not conserve these spectral moments in the resulting Green's function.
The method has a formal $\mathcal{O}[N^6]$ scaling to obtain the full CCSD-GF matrix including off-diagonal elements,
and without any dependence on a grid resolution.
Furthermore,
we show that the approach will also generate an explicit pole representation for the resulting self-energy which characterizes the correlation-driven changes to the underlying mean-field description,
within the appropriate non-Hermitian nature of CC theory.
Finally,
we present results for the approach,
and characterize the resulting method in terms of the convergence of the full-frequency Green's function with respect to number of conserved spectral moments,
and the resulting number of EOM matrix-vector multiplications required.

\subsection{Moment-conserving algorithm}

We first define the precise definition of these occupied (IP) and virtual (EA) GF spectral moments,
which are central to the approach.
Within a formal Hermitian theory,
these are given for an arbitrary state $\bf{\Psi}$ as
\begin{align}
    \label{eq:gf_moment_occ}
    \chi_{pq}^{\mathrm{IP},(m)}
    &=
    \melpsi{\cre_{q} [H_N]^{m} \des_{p}}
    ,
    \\
    \label{eq:gf_moment_vir}
    \chi_{pq}^{\mathrm{EA},(m)}
    &=
    \melpsi{\des_{p} [H_N]^{m} \cre_{q}}
    .
\end{align}
We denote these as `spectral moments',
as they precisely characterize the moments of the individual IP and EA spectral distributions (up to a sign),
as
\begin{align}
    \label{eq:IPmom_dist}
    \chi_{pq}^{\mathrm{IP},(m)}
    &=
    \int_{-\infty}^{\mu} A(\omega)_{pq} \omega^m d\omega
    ,
    \\
    &=
    \sum_{\alpha}
    \langle \Psi | \cre_{q} | \Psi_{\alpha} \rangle
    \langle \Psi_{\alpha} | \des_{p} | \Psi \rangle
    (E_{\alpha}-E_0)^m
    ,
    \\
    \chi_{pq}^{\mathrm{EA},(m)}
    &=
    \int_{\mu}^{\infty} A(\omega)_{pq} \omega^m d\omega
    , \label{eq:EAmom_dist}
    \\
    &=
    \sum_{\alpha}
    \langle \Psi | \des_{p} | \Psi_{\alpha} \rangle
    \langle \Psi_{\alpha} | \cre_{q} | \Psi \rangle
    (E_{\alpha}-E_0)^m
    , \label{eq:EAmom_SOS}
\end{align}
where $\mu$ is the chemical potential,
which can be assumed to be the ground state energy of the $N-$electron system ($E_0$).
$A(\omega)_{pq}$ is the matrix valued spectral distribution,
which can be related to the (time-ordered) Green's function as $A(\omega)_{pq}=-\frac{1}{\pi}  \textrm{Im} [G(\omega)_{pq}]$, or given the separation into hole and particle sectors for these equations, can be equivalently considered as the spectra of the lesser and greater Green's functions for Eq.~\ref{eq:IPmom_dist} and Eq.~\ref{eq:EAmom_dist} respectively. In the sum-over-states representation, $\alpha$ can be considered to sum over $all$ $N\pm1-$electron states with energy $E_{\alpha}$.
Note that these spectral moments include information of all diagonal and off-diagonal elements. 
Furthermore, the relationship between the spectral distribution and Green's function is injective,
given the constraints of the Kramers-Kronig relations between the real and imaginary parts of the Green's function\cite{Fetter2003}.
Due to the bounded nature of the IPs and EAs and rapid decay of the spectral intensity,
the Green's function can be uniquely determined from the spectral moments as the number of given moments increases. The reconstruction of a spectrum from a set of its moments over the intervals of Eqns.~\ref{eq:IPmom_dist} and \ref{eq:EAmom_dist} in this way is the definition of the `Stieltjes moment problem', which posed in 1894 the question of reconstructing probability distributions from its moment expansion, and its uniqueness\cite{Stieltjes1894}.

These IP and EA spectral moments can also be related to the coefficients of a short-time Taylor expansion of the hole and particle propagation from Eq.~\ref{eq:greens_function_time},
as well as the first $2m+1$ terms in the Laurent expansion of the Matsubara Green's function dynamics,
with more detailed given in Ref.~\onlinecite{Fertitta2019}.
Finally,
linear combinations of these moments up to a given order can characterize the coefficients of various orthogonal polynomial expansions of the separate particle or hole Green's function in the real-frequency domain,
such as Chebyshev expansions which have shown to be rapidly convergent \cite{Wolf2014, Dong2020}, connecting this approach to any real-frequency polynomial expansion of the Green's function.

For a mean-field state, all the information on the Green's function is contained in the $m=1$ spectral moment (characterizing the Fock matrix), since all higher moments can be found as simple powers of this. In correlated descriptions, these higher moments are not so simply related, and give rise to the additional spectral features as a result of an effective dynamic self-energy. Specializing to coupled-cluster theory,
these spectral moments corresponding to the EOM-CC Green's function can be written \vtwo{in a rigorous diagrammatic fashion (analogous to Eq.~\ref{eq:greens_function_cc})},
as
\begin{align}
    \label{eq:gf_cc_moment_occ}
    \chi_{pq}^{\mathrm{IP},(m)}
    &=
    \melphi{(1 + \Lambda) \cres_{q} [P_X \bar{H}_N P_X]^{m} \dess_{p}}
    ,
    \\
    \label{eq:gf_cc_moment_vir}
    \chi_{pq}^{\mathrm{EA},(m)}
    &=
    \melphi{(1 + \Lambda) \dess_{p} [P_Y \bar{H}_N P_Y]^{m} \cres_{q}}
    .
\end{align}
The presence of the projection operators in this expression is required to match the standard GFCC approach,
analogous to the truncation of the $X(\omega)$ and $Y(\omega)$ excitation spaces in Eqs.~\ref{eq:cc_x_operator} and \ref{eq:cc_y_operator} to be consistent with the $T$ and $\Lambda$ truncation.
These moments can be simply constructed via recursive application of the standard IP- or EA-EOM $\bar{H}_N$ matrix-vector product on the initial $\cres_{q} | \bf{\Phi} \rangle$ and $\dess_{q} | \mathbf{\Phi} \rangle$ states \cite{Shee2019},
requiring $2N_{\mathrm{orb}}(m-1)$ matrix-vector evaluations to construct all moments up to a desired order $m$.
Note that the $m=0$ spectral moment is simply the coupled-cluster one-body density matrix (or one-hole density matrix),
and does not require such a matrix-vector product, and that all moments (for both IP and EA) span all occupied and virtual indices.

Once these IP and EA spectral moments up to a given (odd) order of $m$ are found,
the construction of an effective subspace Hamiltonian whose Green's function matches these moments can proceed.
The initial aim is to formulate separate tridiagonal subspace Hamiltonians for the occupied and virtual spaces,
conserving their respective IP and EA spectral moments by construction,
with the form
\begin{align}
    \label{eq:tridiagonal_matrix}
    \bar{\mathbf{H}}^\mathrm{IP/EA}_\mathrm{tri}
    =
    \begin{bmatrix}
        \mathbf{A}_0 & \mathbf{C}_1 & \mathbf{0} & \cdots \\
        \mathbf{B}_1 & \mathbf{A}_1 & \mathbf{C}_2 & \cdots \\
        \mathbf{0} & \mathbf{B}_{2} & \mathbf{A}_{2} & \cdots \\
        \vdots & \vdots & \vdots & \ddots 
    \end{bmatrix} ,
\end{align}
where each block is an $N_{\mathrm{orb}} \times N_{\mathrm{orb}}$ matrix.
This block tridiagonal form allows for a straightforward connection to the continued fraction representation of the resulting IP and EA Green's function, via Lanczos-Haydock recursion \cite{Shee2019}.
However,
the traditional non-Hermitian block Lanczos approach to formulating this matrix is fundamentally changed,
such that the recursion is centered around conservation of the resulting spectral moments,
rather than just defining a Krylov subspace of higher powers of $\bar{H}_N$ \cite{Weikert1996}.
This is adapted from the Hermitian formulation of Ref.~\onlinecite{Sriluckshmy2021} and fundamental work on the Stieltjes moment problem to work with the non-Hermitian moments of CC theory\cite{Derkach2020, Schmdgen2017}. 

First,
the moments are orthogonalised under the metric of the zeroth moment (transforming to the natural orbital basis),
independently for the IP and EA sectors (where explicit indices are dropped in subsequent equations for clarity,
with all quantities considered as $N_{\mathrm{orb}} \times N_{\mathrm{orb}}$ matrices for both the IP and EA sectors),
as
\begin{align}
    \mathbf{S}_{m}
    &=
    (\boldsymbol{\chi}^{(0)})^{-\frac{1}{2}}
    \boldsymbol{\chi}^{(m)}
    (\boldsymbol{\chi}^{(0)})^{-\frac{1}{2}}
    .
\end{align}
Following this,
the elements of Eq.~\ref{eq:tridiagonal_matrix} are constructed according to the recursive formulae
\begin{align}
    \label{eq:block_lanczos_b2}
    \mathbf{B}^{2}_{n+1}
    &=
    \sum_{j=0}^{n+1}
    \sum_{l=0}^{n}
    \mathbf{W}^{n}_{l} \mathbf{S}_{j+l+1} \mathbf{V}^{n}_{j-1}
    -
    \mathbf{A}^{2}_{n}
    -
    \mathbf{C}^{2}_{n},
    \\
    \label{eq:block_lanczos_c2}
    \mathbf{C}^{2}_{n+1}
    &=
    \sum_{j=0}^{n+1}
    \sum_{l=0}^{n}
    \mathbf{W}^{n}_{j-1} \mathbf{S}_{j+l+1} \mathbf{V}^{n}_{l}
    -
    \mathbf{A}^{2}_{n}
    -
    \mathbf{B}^{2}_{n},
    \\
    \label{eq:block_lanczos_v}
    \mathbf{V}^{n+1}_{j}
    &=
    [
        \mathbf{V}^{n}_{j-1}
        -
        \mathbf{V}^{n}_{j} \mathbf{A}_{n}
        -
        \mathbf{V}^{n-1}_{j} \mathbf{B}_{n}
    ]
    \mathbf{C}^{-1}_{n+1}
    ,
    \\
    \label{eq:block_lanczos_w}
    \mathbf{W}^{n+1}_{j}
    &=
    \mathbf{B}^{-1}_{n+1}
    [
        \mathbf{W}^{n}_{j-1}
        -
        \mathbf{A}_{n} \mathbf{W}^{n}_{j}
        -
        \mathbf{C}_{n} \mathbf{W}^{n-1}_{j}
    ]
    ,
    \\
    \label{eq:block_lanczos_a}
    \mathbf{A}_{n+1}
    &=
    \sum_{j=0}^{n+1}
    \sum_{l=0}^{n+1}
    \mathbf{W}^{n+1}_{l} \mathbf{S}_{j+l+1} \mathbf{V}^{n+1}_{j}
    ,
\end{align}
where the coefficients $\mathbf{V}$ and $\mathbf{W}$ satisfy
\begin{align}
    \label{eq:block_lanczos_v_satisfies}
    \ket{v_n}
    &=
    \sum_{j=0}^{n}
    H^j
    \ket{v_0}
    \mathbf{V}_{j}^{n}
    ,
    \\
    \label{eq:block_lanczos_w_satisfies}
    \bra{w_n}
    &=
    \sum_{j=0}^{n}
    \mathbf{W}_{j}^{n}
    \bra{w_0}
    H^j
    .
\end{align}
It must therefore follow that the initial coefficients are are identity,
\begin{align}
    \label{eq:block_lanczos_vw_initial}
    \mathbf{V}_0^0
    &=
    \mathbf{W}_0^0
    =
    \mathbf{I}
    .
\end{align}
Traversing the recursive formulae in Eq.~\ref{eq:block_lanczos_b2}-\ref{eq:block_lanczos_vw_initial} results in a maximum number of $\frac{1}{2} (m+1)$ $\mathbf{A}$ blocks.
The coarse-grained excitation energies of each IP and EA sector can therefore be obtained via eigenvalue decomposition of Eq.~\ref{eq:tridiagonal_matrix}
\begin{align}
    \label{eq:tridiagonal_eig}
    \bar{\mathbf{H}}_\mathrm{tri}
    &=
    \tilde{U}
    E
    \tilde{U}^{-1}
    ,
\end{align}
where the first $N_\mathrm{orb}$ rows of eigenvectors $\tilde{U}$ can be back-transformed from the orthogonal basis to give left- and right-hand transition amplitudes,
corresponding to the weight of each excitation over the physical degrees of freedom, as
\begin{align}
    \label{eq:rotate_u}
    U
    &=
    [\boldsymbol{\chi}^{(0)}]^\frac{1}{2}
    P_\mathrm{phys}
    \tilde{U}
    ,
    \\
    \label{eq:rotate_v}
    V
    &=
    (
    \tilde{U}^{-1}
    P_\mathrm{phys}
    [\boldsymbol{\chi}^{(0)}]^\frac{1}{2}
    )^\dagger
    .
\end{align}
The projection into the physical space is defined as
\begin{align}
    P_\mathrm{phys}
    &=
    \sum%
    _{p \in \mathrm{phys}}%
    ^{N_\mathrm{orb}}
    \ket{p} \bra{p}
\end{align}
where $p$ enumerate physical states,
represented by molecular orbitals at the Hartree--Fock level.
The energies, $E$, and vectors, $\{U,V\}$, of these states are then sufficient to recover the original moments of the Green's function, as
\begin{align}
    \label{eq:compressed_moments}
    \chi^{(m)}_{pq}
    &=
    \sum_{\alpha}
    U_{p\alpha}
    E_{\alpha}^m
    V_{q\alpha}^\dagger
    ,
\end{align}
where $\alpha$ now runs over excitations resulting from both the hole and particle Hamiltonian construction, and $p,q$ run over the `physical' space of MOs. Critically, these spectral moments will exactly reproduce the original CC spectral moments of Eqns.~\ref{eq:gf_cc_moment_occ} and \ref{eq:gf_cc_moment_vir} provided, by construction to numerical precision. However, we can now also find a full spectrum on an arbitrary grid of frequency points, for which these moments provide the constraints, as
\begin{align}
    \label{eq:compressed_spectrum}
    A_{pq}(\omega)
    &=
    -\frac{1}{\pi}
    \sum_{\alpha}
    \mathrm{Im}
    \bigg[
    \frac{U_{p\alpha} V_{q\alpha}^\dagger}{\omega - E_\alpha - i\eta}
    \bigg]
    .
\end{align}
The number of poles in the approximate spectrum (indexed by $\alpha$ in Eq.~\ref{eq:compressed_spectrum}, spanning both occupied and virtual states) will grow as the number of provided moments increases, as (at most) $N_{\rm orb} (m+1)$, compared to the number of mean-field excitations of simply $N_{\rm orb}$.
These additional states describe the correlation-induced splitting and adjustment of spectral intensity, emergence of quasiparticle lifetimes, and introduction of satellite peaks in the spectrum. Since each recursion of the modified block Lanczos iterations requires two additional GF moments as input, we define the approach via an integer $n$, which conserved all moments $0\leq m \leq 2n+1$, giving the heirarchy of approximations we denote GF$(n)$. \vtwo{While this final `moment-constrained' Green's function can not be simply written in terms of a connected diagrammatic expansion, the moments of its spectral distribution will be rigorously diagrammatic by construction, which will therefore tend to the exact diagrammatic Green's function of coupled-cluster theory as $n$ is increased.}

This algorithm \vtwo{therefore} provides a systematic series of approximations,
with inclusion of each additional order of $n$ requiring further computational effort, but yielding a Green's function which is closer to the exact dynamic limit. An increase of $n$ by one provides an additional (up to) $2N_\mathrm{orb}$ poles in the quasiparticle spectrum, and requires an additional two moments in each sector.
The algorithm presented scales only cubically with system size (once the spectral moments have been computed, which is itself \vtwo{$\mathcal{O}[N_{\mathrm{orb}}(N_{\mathrm{occ}}^4 N_{\mathrm{vir}} + N_{\mathrm{occ}} N_{\mathrm{vir}}^4)]$ per moment} at the level of CCSD),
but is capable of producing a spectrum of arbitrary resolution in frequency space.
This is a significant improvement upon conventional algorithms which may require multiple evaluations of an $\mathcal{O}[N^6]$ step at each frequency point of interest,
constraining the frequency-resolution by available computational resources. However, if very accurate resolution of high-frequency dynamics at the level of CCSD is required, then the moment expansion may not be the most efficient approach compared to a direct frequency-domain targeting algorithm \cite{Nooijen1992, Nooijen1993, Peng2019}, as very high moments may start to be beset by numerical difficulties, which likely require further developments (e.g. reorthogonalization steps) to resolve.

\subsection{Explicit self-energy construction}

One may also wish to algebraically cast the excitations of this spectrum back into a set of static `auxiliary' degrees of freedom which,
when folded down into frequency space over `physical' degrees of freedom (e.g. the MO space),
represent the effect of a dynamic self-energy \cite{Lowdin1962, Hirata2015, Hirata2017, Loos2020, Backhouse2021, Bintrim2021}. This gives a pole representation of the self-energy required to achieve the spectrum of Eq.~\ref{eq:compressed_spectrum}, without resorting to an explicit Dyson equation requiring a potentially ill-conditioned and costly inversion of the Green's function at each frequency.
To do this, we need to combine the particle and hole excitations into a single hamiltonian, such that its diagonalization gives the eigenenergies given by $E_{\alpha}$, and where the projections of the corresponding eigenvectors onto the physical space give back $U$ and $V$. The self-energy can then be represented as the part of this Hamiltonian external to the physical space (describing the augmentation of the MO space due to the self-energy).

First we define vectors spanning the physical space consisting of the projections of the excitations on the MO space, concatenating the vectors for both the IPs and EAs of Eqns.~\ref{eq:rotate_u}-\ref{eq:rotate_v}, as
\begin{align}
    \label{eq:combined_vector_left}
    U
    &=
    \begin{bmatrix}
        U_h & U_p
    \end{bmatrix}
    ,
    \\
    \label{eq:combined_vector_right}
    V
    &=
    \begin{bmatrix}
        V_h & V_p
    \end{bmatrix}
    ,
\end{align}
which have dimension $N_\mathrm{orb} \times (L + M)$,
where $L$ and $M$ are the sizes of the compressed IP and EA spaces,
respectively.
One can then construct an additional $L + M - N_\mathrm{orb}$ rows of these matrices, in order to define augmented (square), full-rank $U_c$ and $V_c$ matrices of dimension $(L + M) \times (L + M)$, while maintaining their projection onto the initial $N_\mathrm{orb} \times N_\mathrm{orb}$ physical subspace as given by Eqns.~\ref{eq:combined_vector_left}-\ref{eq:combined_vector_right}. This can be achieved using any complete biorthogonal basis which does not change the existing vectors in the physical space, e.g. via a two-sided Gram-Schmidt,
or using the eigenvectors corresponding to the non-null-space of $\mathbf{I} - U^\dagger V$. These vectors and the IP and EA excitation energies can then form a biorthogonal eigenbasis for a new hamiltonian with the desired properties, spanning the `physical' MO space coupled to an external space, which has exactly the spectrum of Eq.~\ref{eq:compressed_spectrum} when projected back into the physical space. We can define this Hamiltonian as
\begin{align}
    \label{eq:combined_compressed_hbar}
    \bar{\mathbf{H}}_\mathrm{comb}
    &=
    U_c
    \begin{bmatrix}
        E_p & 0 \\ 0 & E_h
    \end{bmatrix}
    V_c^\dagger
    .
\end{align}
In order to obtain an explicit pole representation of this effective self-energy, $\bar{\mathbf{H}}_\mathrm{comb}$ must be rotated into a representation in which the non-physical (`auxiliary') subspace is diagonal, using the eigenvectors of $P_\mathrm{ext} \bar{\mathbf{H}}_\mathrm{comb} P_\mathrm{ext}$,
where $P_\mathrm{ext} = 1 - P_\mathrm{phys}$, and resulting in the ($L+M \times L+M)$ matrix, $\bar{\mathbf{H}}_\mathrm{aux}$.
The energies of the poles of the resulting effective self-energy are denoted $\epsilon$, obtained from the diagonal of $P_\mathrm{ext} \bar{\mathbf{H}}_\mathrm{aux} P_\mathrm{ext}$,
with the left- and right-hand residues as the vectors $\lambda = P_\mathrm{phys} \bar{\mathbf{H}}_\mathrm{aux} P_\mathrm{ext}$ and $\mu^\dagger = P_\mathrm{ext} \bar{\mathbf{H}}_\mathrm{aux} P_\mathrm{phys}$,
respectively. This allows the explicit self-energy to be written as
\begin{align}
    \mathbf{\Sigma}(\omega) = (P_\mathrm{phys} \bar{\mathbf{H}}_\mathrm{aux} P_\mathrm{phys} - \mathbf{F}) + \sum_{\kappa}^{L+M-N_\mathrm{orb}} \lambda_{\kappa} \frac{1}{\omega-\epsilon_{\kappa}} \mu_{\kappa}^\dagger, \label{eq:self-energy}
\end{align}
where $F$ is the original Fock matrix in the physical space, and $\kappa$ runs over the states in the external space.
Similarly, the Green's function can then be written via a Dyson equation as
\begin{align}
    \label{eq:combined_gf}
    \mathbf{G}(\omega)
    &=
    \bigg(
    \omega \mathbf{I}
    -
    P_\mathrm{phys} \bar{\mathbf{H}}_\mathrm{aux} P_\mathrm{phys}
    -
    \sum_{\kappa}^{L+M-N_\mathrm{orb}} \lambda_{\kappa}
    \frac{1}{\omega - \epsilon_{\kappa}}
    \mu_{\kappa}^\dagger
    \bigg)^{-1}
    ,
\end{align}
where the spectrum of Eq.~\ref{eq:combined_gf} above will match that of Eq.~\ref{eq:compressed_spectrum}, with the distinction being that it arises from the diagonalization of a single Hamiltonian, allowing for an explicit self-energy to extracted, rather than from a combination of separate particle and hole Hamiltonians. 


\vtwo{One advantage of this explicit self-energy construction is that Fermi liquid parameters such as effective masses and renormalization factors of individual states can be found, which characterize the magnitude of the correlation-driven changes to low-energy excitations, despite not being formal observables themselves. While in condensed matter they can be used to define interaction-driven quantum phase transitions, in molecular and finite systems they can act as a useful proxy for defining the effect of correlations on a specific (molecular) energy level. This quasiparticle renormalization factor for a given MO, $i$, can be written as
\begin{equation}
    Z_i = \left( 1 - \left. \frac{\partial \Sigma_{ii}(\omega)}{\partial \omega} \right|_{\omega=0} \right)^{-1}.
\end{equation}
When the explicit form of Eq.~\ref{eq:self-energy} is considered, this reduces to
\begin{equation}
    Z_i = \left( 1 + \sum_{\kappa}^{L+M-N_\textrm{orb}} \frac{\lambda_{i \kappa} \mu^*_{i, \kappa}}{\epsilon_{\kappa}^2} \right)^{-1} .  \label{eq:QP_renorm}
\end{equation}
This renormalization factor, $Z_i$, takes values between zero and one, where one indicates that the correlations have not changed the state at all from its original mean-field description (Hartree--Fock), while smaller values indicate that the stronger correlations have increasingly shifted the energy of the state. We note that for finite systems, this renormalization factor loses its precise motivation in terms of the changing character of electronic structure at the Fermi surface of a metal. However, as a heuristic for the strength of correlation-driven changes to the frontier excitations about the chemical potential, it is still a valid diagnostic. To characterize correlation-driven changes to higher-energy states away from the chemical potential, the derivative of the self-energy at different energies (i.e. the Fock MO energy of the orbital in question) can be used instead.
}

\subsection{Technical details}

We have implemented the present algorithm in the \texttt{PySCF} package \cite{PySCF2017, PySCF2020},
where we look to incorporate it as public-accessible functionality in the near future.
The code supports MPI parallelism,
distributing the work required to compute the matrix-vector product between MPI processes.
Within each MPI process, the matrix-vector product is builds on the existing \texttt{PySCF} EOM-CC functionality \cite{PySCF2017, PySCF2020, Wang2020},
which already supports OpenMP parallelism,
and we therefore achieve a hybrid parallel algorithm, where each MPI rank is a separate physical node, and OpenMP communication within the nodes, resulting in an effective implementation in high-performance computing settings whilst retaining a lightweight codebase.
Our order of operations is designed to reduce communication of large arrays, whereby a job with $K$ available MPI processes (generally nodes) proceeds as
\begin{enumerate}
    \item Compute hole right-hand side
    $
        \dess_q
        \ket{\boldsymbol{\Phi}}
    $
    and particle right-hand side
    $
        \cres_q
        \ket{\boldsymbol{\Phi}}
    $
    on all MPI processes.
    \item Compute hole left-hand side
    $
        \bra{\boldsymbol{\Phi}}
        (1 + \Lambda)
        \cres_p
        [P_X \bar{H}_N P_X]^m
    $
    and particle left-hand side
    $
        \bra{\boldsymbol{\Phi}}
        (1 + \Lambda)
        \dess_p
        [P_Y \bar{H}_N P_Y]^m
    $ up to desired moment order, where the effort is distributed amongst the MO indices of $p$, assigned to process $(p~\mathrm{mod}~K)$.
    \item Contract left- and right-hand sides for hole and particle cases according to Eqns.~\ref{eq:gf_cc_moment_occ} and \ref{eq:gf_cc_moment_vir}
    for elements $\chi^{(m)}_{pq}$ on  process $(p~\mathrm{mod}~K)$.
    \item Perform a reduction operation to gather $\chi$ between all processes.
\end{enumerate}
This algorithm is therefore embarrassingly parallel for the computation of the moments, assuming the ability to perfectly distribute the load over the MO indices on each MPI rank. 
With this algorithm we have been able to perform calculations on the full GW100 benchmark set in a def2-TZVPP basis set (see sec.~\ref{sec:GW100}),
with the largest system in this set having 411 orbitals and 78 correlated electrons.
The total run time for the calculation of GF(0) through GF(5) for this system was approximately 2 hours on 32 nodes, where each node (consisting of two AMD 7742, 2.25 GHz, 64-core CPUs) was assigned a separate MPI process.

%
%

\section{Results and discussion} \label{sec:results}

\subsection{Proof of principle}

\begin{table}[tb!]
    \centering
    \includegraphics[width=0.49\textwidth]{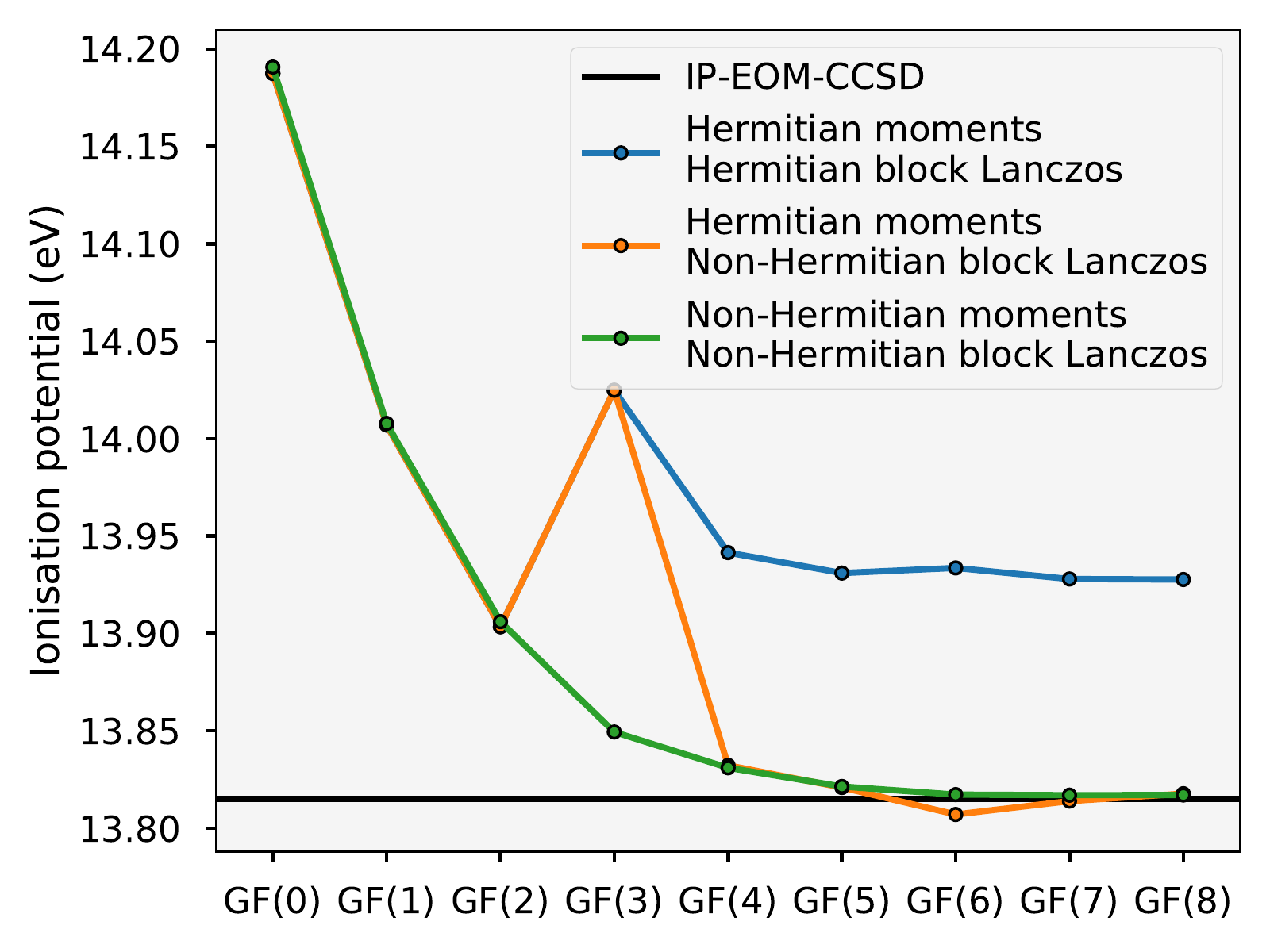}%
    \captionof{figure}{
        Comparison of the convergence in the ionisation potential for a CO molecule (bond length of 1.1314 $\mathrm{\AA}$) in a cc-pVDZ basis between the Hermitian and non-Hermitian block Lanczos recursion, showing the correct convergence of the non-Hermitian form to the exact IP-EOM-CCSD ionization potential, and incorrect convergence or numerical errors for Hermitized versions. Note that the HOMO energy at the level of the Hartree--Fock starting point is 19.12 eV, showing that even at the lowest levels of moment conservation, significant correlation-driven changes to the IP are found.
    }
    \label{fig:moment_convergence}
\end{table}
In Fig.~\ref{fig:moment_convergence}, we show convergence of the (first) ionisation potential of Carbon Monoxide in a cc-pVDZ basis, with increasing orders of CCSD Green's function spectral moment conservation. We stress again that the approach does not directly target this IP, which is more common to `state-specific' approaches to computing excited states in e.g. ADC or EOM methods \cite{Schirmer1983, Stanton1993}, but instead the method involves satisfaction of moment constraints which are integrated quantities over all excitations in the spectrum (Eqns.~\ref{eq:IPmom_dist}-\ref{eq:EAmom_SOS}), and therefore convergence of any single excitation is expected to be a stern test. Nevertheless, the aim is to systematically converge to all excitations in the IP/EA-EOM-CCSD spectrum. 

We also compare three different variants of the algorithm; the one outlined in Sec.~\ref{sec:MomCCGF}, one where we modify each moment to the nearest Hermitian form (in a least squares sense) before computation of the resulting spectrum (as $\frac{1}{2} (\mathbf{\chi}^{(m)} + \mathbf{\chi}^{(m) \dagger})$), and one where the moments are Hermitized {\em and} the algorithm of Sec.~\ref{sec:MomCCGF} is constrained to be Hermitian (i.e. $\mathbf{B}_i=\mathbf{C}_i^{\dagger}$ in Eq.~\ref{eq:tridiagonal_matrix}). Note that the emergence of non-Hermitian moments is a feature of the non-unitary nature of CC theory used in the moment construction, with other levels of theory (and the `exact' moments) expected to be Hermitian by construction. 
The use of a Hermitian (real-valued) block Lanczos recursion of Eqns.~\ref{eq:block_lanczos_b2}-\ref{eq:block_lanczos_a} requires $\mathbf{B}^2$ to be positively-semi-defined in order to compute $\mathbf{B}$, and positively-defined to compute $\mathbf{B}^{-1}$.
Any space corresponding to negative (or zero) eigenvalues in $\mathbf{B}^2$ must therefore be discarded,
resulting in the possibility of loss of flexibility in the resulting effective Hamiltonian.
In the non-Hermitian recursion,
$\mathbf{B}$ and indeed $\mathbf{C}$ are allowed to be complex and therefore no such states need be discarded,
except any null space when inverting the matrices.

The error due to the approximation of Hermiticity in the EOM-CCSD moments themselves, 
whilst still allowing $\mathbf{B}$, $\mathbf{B}^{-1}$, $\mathbf{C}$ and $\mathbf{C}^{-1}$ to be complex,
does not produce errors as large in magnitude as the real-valued constraint.
This is in agreement with the claim of Shee and Zgid that the error between the exact and symmetric part of the coupled cluster Green's function is generally small\cite{Shee2019}.
Despite this,
we can observe that the convergence of the excitations in the case of the complex-valued recursion and Hermitised moments is less rigorous,
and there is more sensitivity to numerical error.
Furthermore,
manual Hermitisation of the moments does not necessarily result in a matrix that is guaranteed to have the correct definite structure and causal Green's function and self-energy (again, a feature of CC theory).
We note that Fig.~\ref{fig:moment_convergence} is an example selected where the ill behaviour of the Hermitised moments is particularly large,
and not all systems have such a large discrepancy between the Hermitian and non-Hermitian variants. Nevertheless, we can clearly demonstrate systematic convergence to the IP-EOM-CCSD ionization potential with the appropriate non-Hermitian recursions.

\subsection{Full spectrum convergence}

\begin{figure}[htb!]
    \centering
    \subfloat[
        \centering
        1.1 $\mathrm{\AA}$
    ] {
        \includegraphics[width=0.49\textwidth]{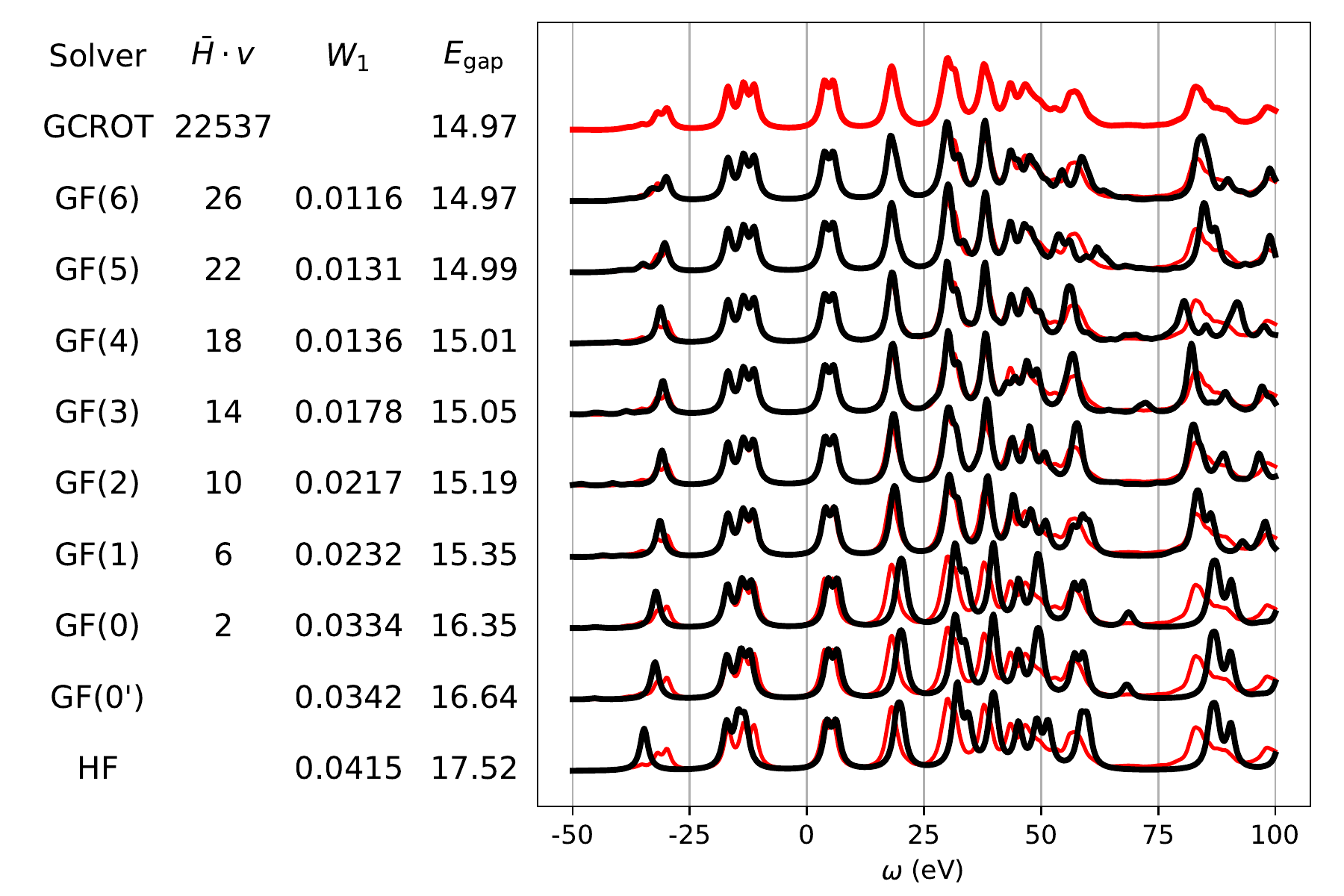}
    }
    \subfloat[
        \centering
        1.8 $\mathrm{\AA}$
    ] {
        \includegraphics[width=0.49\textwidth]{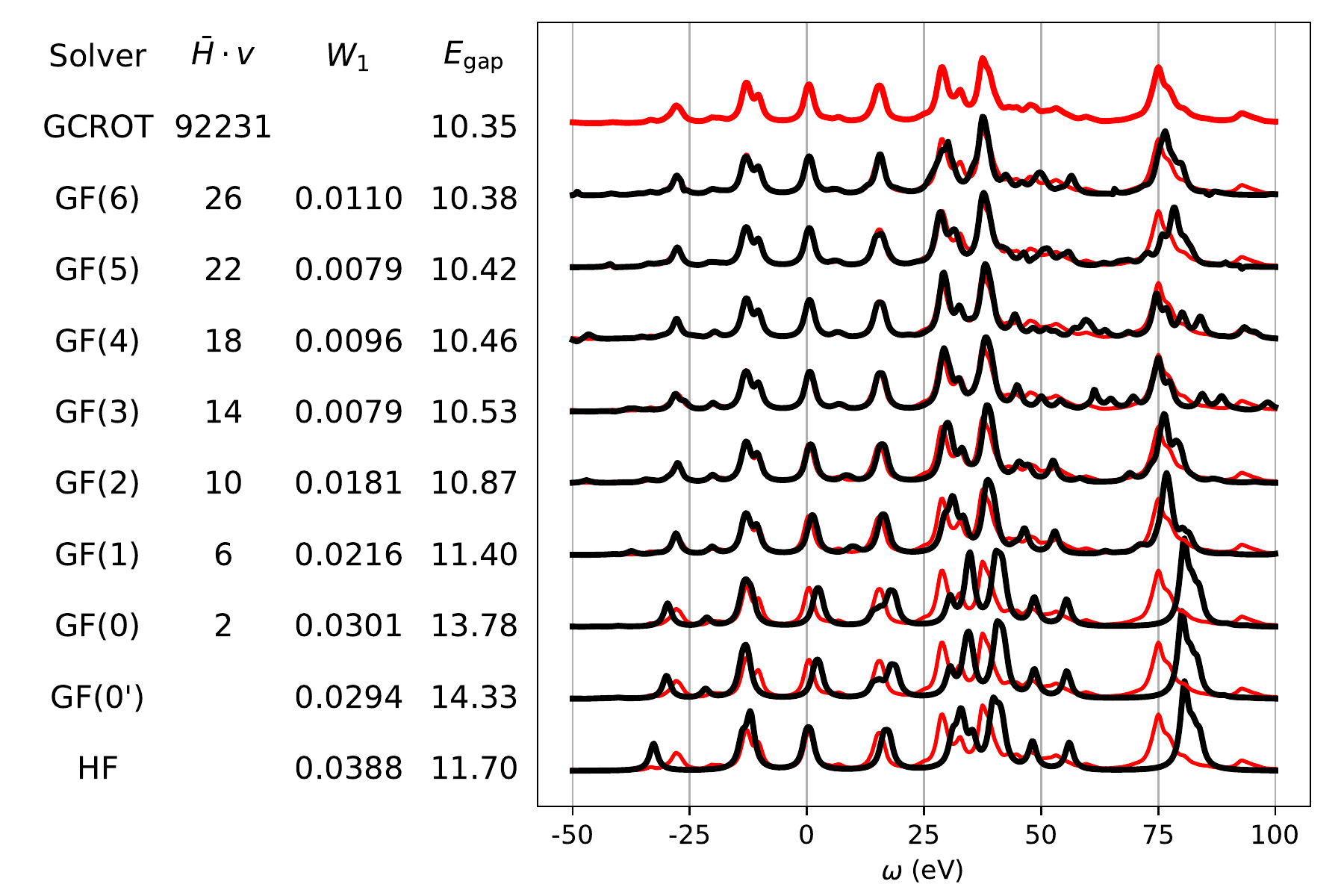}
    }
    \caption{
        Comparison of theoretical photoelectron spectra computed using the recursive GF moment conservation (presented in Sec.~\ref{sec:MomCCGF} in black),
        and a more conventional GF-CCSD approach using the GCROT algorithm (in red under-laying all spectra),
        for water with a bond length of (a) 1.1 $\mathrm{\AA}$ and (b) 1.8 $\mathrm{\AA}$ in a cc-pVDZ basis.
        The notation GF($n$) indicates the number of iterations of the recursion algorithm performed, conserving all GF moments up to order $0 \leq m \leq 2n+1$.
        GF($0^\prime$) corresponds to a modified GF(0) approximation where the moments are exactly computed via the reduced density matrices (Sec.~\ref{sec:GF_prime}).
        Also shown is the spectrum at the mean--field (Hartre--Fock) level.
        The labels also give the number of matrix-vector products per orbitals required to produce the spectrum,
        where the GCROT result depends on the number of frequency points $N_\omega$,
        which was selected to be 512 in this example with a broadening parameter $\eta$ of $1.0~\mathrm{eV}$ (the other GF results are artificially broadened with the same broadening).
        The Wasserstein metric $W_1$ between each of the spectra and the GCROT spectrum is shown,
        indicating the their fit to the true GF-CCSD spectrum.
        The value of the gap for each method is also included, with the chemical potential at the zero frequency.
    }
    \label{fig:compare_to_dynamic}
\end{figure}
The strength of the method (over e.g. an EOM approach) will likely be found in the fact that the entire excitation spectrum can be converged (on the real-frequency axis) to a good accuracy simultaneously, with arbitrarily small broadening. To demonstrate this, in Fig.~\ref{fig:compare_to_dynamic} we show the convergence of the full spectrum with increasing numbers of moments,
to a more conventional `correction-vector' implementation of CCGF as outlined in Sec.~\ref{sec:CCGF} using the GCROT algorithm to solve the linear equations of Eqns.~\ref{eq:cc_x_relation}-\ref{eq:cc_y_relation} at each frequency point of interest and with a fixed broadening ($\eta$) of the line shapes\cite{deSturler1999, Parks2006, Hicken2010, SciPy2020}.
A water molecule in a cc-pVDZ basis was used with a total of 24 orbitals,
in both an equilibrium and stretched (more strongly correlated) configuration.
The number associated with the labels gives the required number of matrix-vector computations per orbital between $\bar{H}$ and an arbitrary state vector, which dominates the computational cost.
This number scales with the number of frequency points $N_\omega$ only in the case of the conventional (GCROT) CCGF calculation,
with $N_\omega=512$ used in this plot and broadening to regularize the linear equations of 1.0 eV.
The number of required matrix-vector products for a conventional calculation is typically considerably larger than our modified block Lanczos recursion approach to building a representative Hamiltonian from the GF moments, with all `moment-conserving' spectra shown computed for less than the average cost of a {\em single} frequency point in the traditional frequency-domain approach.

The low-energy excitations around the chemical potential are found to rapidly converge as the numbers of moments conserved increases, with the quasiparticle gap ($E_\mathrm{gap}$) shown to converge to less than 0.1 eV error compared to the `exact' EOM-CCSD one by GF(4) in both correlation regimes (corroborated more broadly by the test set of Sec.~\ref{sec:GW100}, and which is roughly the intrinsic accuracy of the CCSD method). The only anomalous point is the GF(0) result in the more strongly correlated regime, where the lack of spectral information in the first two hole and particle moments results in a gap which is in error by even more than the initial Hartree--Fock estimate. \vtwo{The importance of correlations in modifying the Hartree--Fock energy levels in the stretched case compared to the equilibrium molecule can be quantified by considering the quasiparticle renormalization factor for the HOMO state (see Eq.~\ref{eq:QP_renorm}). For the equilibrium case, $Z_{\mathrm{HOMO}}=0.93$, while in the stretched case it is $Z_{\mathrm{HOMO}}=0.27$, indicating a signficant a qualitative change in the HOMO state upon stretching of the bonds.}
Individual higher-energy excitations are found to converge more slowly, despite the broad trends of spectral density over higher energies being well reproduced also by GF(4). To quantify this accuracy over the full spectral range, we also provide the Wasserstein (or `earthmover') metric ($W_1$) between each spectrum and the exact CCGF one, which appropriately characterizes the shift in overall spectral weight between two probability distributions. This is found to converge in an almost entirely monotonic fashion with the number of conserved moments, demonstrating the systematic convergence over the whole spectral range. 

A considerable advantage of the GF moment approach is that an explicit pole structure of both the overall approximation to the GF and self-energy is produced, allowing one to adjust the plotting parameters such as the number of frequency points $N_\omega$,
and the broadening parameters $\eta$.
A traditional CCGF correction vector algorithm requires one to select these parameters beforehand and their adjustment requires new calculations.
We note that correction vector algorithms are also more difficult to converge around the excitation energies, requiring substantially more iterations as the condition number becomes larger and the linear equations of Eqns.~\ref{eq:cc_x_relation}-\ref{eq:cc_y_relation} become increasingly singular. While these are not problems in the present algorithm, numerical difficulties exist in a different limit.
In exact-precision arithmetic the limit of large numbers of moments should exactly reproduce the exact CCGF result (formally the complete limit for CCSD will scale as $\mathcal{O}[N^2]$ moments). However, floating-point arithmetic makes very large moments difficult to work with,
as successively higher powers of $\bar{H}$ are required.
Additionally,
a loss in biorthogonality between the left- and right-hand Lanczos block vectors is observed at high moment numbers,
which may introduce further errors within finite-precision arithmetic.
We consider the improvement of the numerical accuracy and maintenance of biorthogonality to be out of the scope of this current work, where we do not extend to very high moment numbers,
and future work will look to remedy this.

\begin{figure}[htb!]
    \centering
    \includegraphics[width=0.98\textwidth]{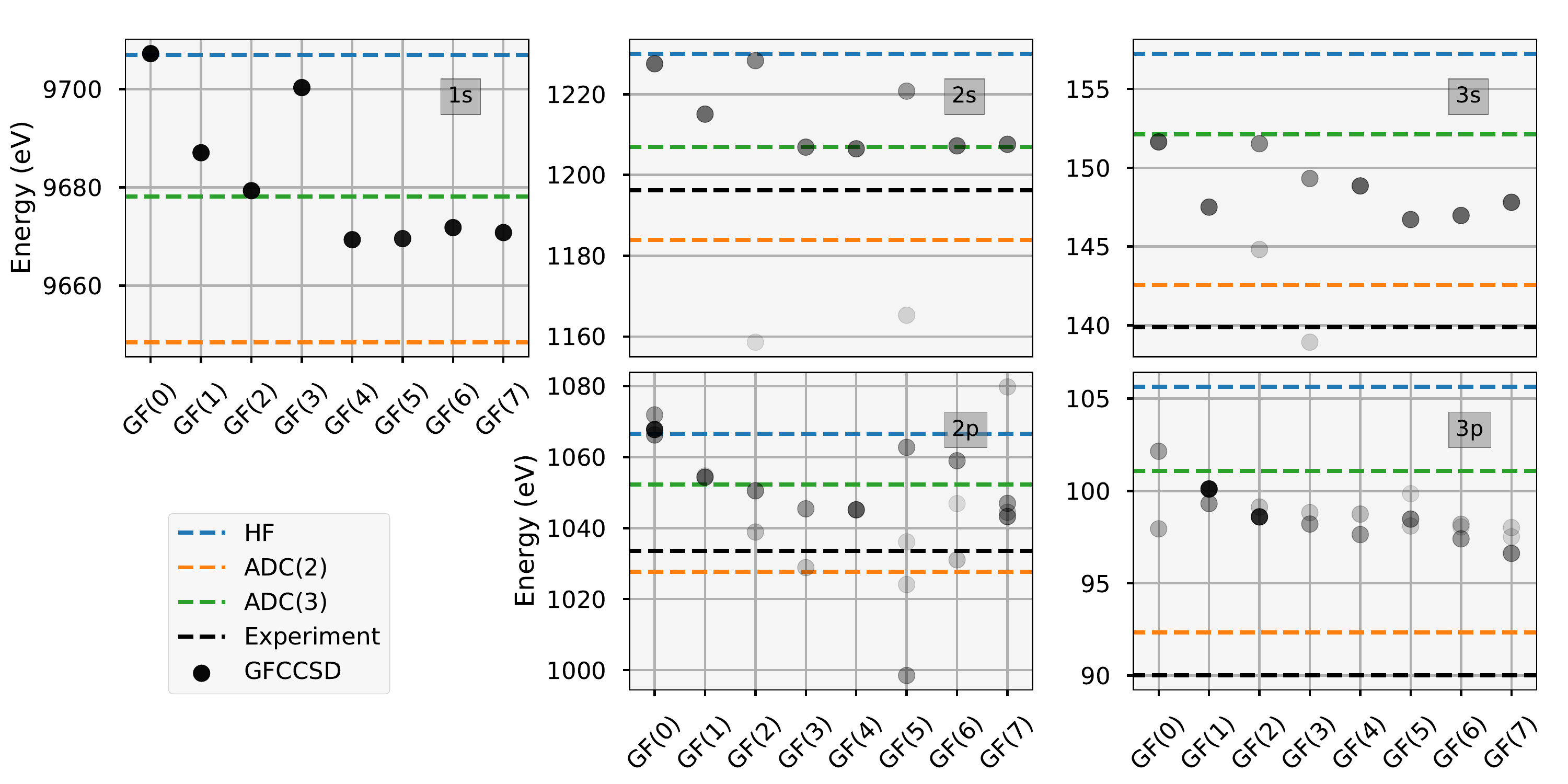}
    \caption{
        \vtwo{Convergence of CCSD core-level ionization energies from $K-$ ($1s$), $L-$ ($2s$, $2p$) and $M-$ ($3s$, $3p$) shells of a Zinc atom with increasing moment constraints, in a cc-pwCVTZ basis with X2C Hamiltonian. The transparency in the shading of points is proportional to the weight of the resulting ionization potential on the atomic orbital of the same character as the desired excitation. This shows that different moment constraints can sometimes result in a splitting of the state across a range of energies.
        The ADC(2) and ADC(3) results are taken from Ref.~\cite{doi:10.1063/1.5131771} and experimental results from Ref.~\cite{PhysRevB.8.2392}. We note that experimental values for the $2p$ and $3p$ orbitals are obtained by averaging ionization energies for states with the total angular momentum quantum number $J = 1/2$ and $J = 3/2$.}}
    \label{fig:core_excits}
\end{figure}

\vtwo{One area where high-accuracy prediction of energy levels is increasingly important is core-level spectroscopy. Similar to Fig.~\ref{fig:compare_to_dynamic}, we will see that the convergence of core levels with respect to moment expansion is slower and less reliable than the frontier states about the chemical potential, which also can suffer from loss of numerical accuracy and maintenance of biorthogonality at higher moments. In Fig.~\ref{fig:core_excits} we consider the core energy levels from the $K-$, $L-$ and $M-$shells of a Zinc atom in a large cc-pwCVTZ basis\cite{corebasis}, with a two-component X2C relativistic Hamiltonian\cite{relhams}. We compare to experimental results of Ref.~\onlinecite{PhysRevB.8.2392}, and recent ADC(2) and ADC(3) calculations \cite{doi:10.1063/1.5131771}, without making the common core-valence separation approximation \cite{Cederbaum_1980,PhysRevA.22.206}. We also show the energy of all relevant ionization potentials, where the shading indicates the weight of the excitation on the atomic orbital with the same character. It can be seen that while sometimes the excitation level is clearly defined and quasiparticle-like, with the state being dominated by a single atomic orbital, for other cases we find that the state can be split into multiple states over a range of relevant energies, with lower weights on the atomic orbital of relevance.}

\vtwo{This splitting of excitations could be physical, driven by the correlations inducing an effective broadening of the quasiparticle lifetimes and other satellite features, or it could be an artifact from not fully converging with respect to moment order, or it could be a numerical feature of finite precision arithmetic. It should be stressed that the moments of the final Green's functions match the original moments provided, with these deep energy-scales not directly probed by the method. Rather, these energy levels have to be resolved via constraining the moments of the overall spectrum at all energy scales, and there are often multiple ways which this can be satisfied. However, at the higher moment orders it does seem as though reasonable convergence and accuracy of these core states is achievable, with the scatter of the larger-weighted states generally below the difference in energy in going from an ADC(2) to ADC(3) level of theory. Despite this, the precise energy of these core states is not fully converged with respect to moment order in all cases, which is exacerbated by the numerical uncertainties which can result from these high moment orders and wide energy scales arising from low-energy states and large basis sets. Further improvements will aim to improve this stability to go to higher moment orders in the future.}

\subsection{A `ground state' approximation to the spectrum} \label{sec:GF_prime}

In Fig.~\ref{fig:compare_to_dynamic}, we also show an alternate approximation denoted GF($0^\prime$). In this, the first two particle and hole CCSD GF moments ($m=0$ and $1$) are computed not from an EOM approach of Eqns.~\ref{eq:gf_cc_moment_occ}-\ref{eq:gf_cc_moment_vir}, but instead are found directly from the one- and two-body reduced density matrices (RDMs) of the ground state, which are readily available from many codes (and indeed from different levels of theory). Appendix A details how the $m=1$ moment can be derived from these RDMs, and how the `commutator trick' effects the rank reduction required for their computation. This derivation is rigorous for variational methods, and will only approximately hold for coupled-cluster theory.
Nevertheless, this shows how a limited level of information about the spectrum of charged particle fluctuations can be obtained from these RDMs. This approximation is however fundamentally different at the CCSD level of theory compared to the GF(0) approximation, since the GF moments of Eqns.~\ref{eq:gf_cc_moment_occ}-\ref{eq:gf_cc_moment_vir} are computed via RDMs in the \emph{absence} of the projection operators ($P_X$ and $P_Y$) onto the $2p1h$ and $2h1p$ spaces that the CC Hamiltonian is represented in. In this way, computing the moments from the RDMs allows for limited fluctuations into the $3p2h$ and $3h2p$ spaces which are absent in a CCSD treatment. 

The consequence of this is that the spectrum from the GF($0^\prime$) approximation has some favourable properties. Chiefly, these are that the energy computed from the spectrum will be equal to the CCSD energy via the Migdal-Galitskii formula \cite{Fertitta2019} -- a limitation of traditional CCSD Green's functions which has been noted from early in their formulation \cite{Nooijen1992,Nooijen1993}. However, despite allowing for a greater flexibility in the description of the fluctuation space of the excitations, this doesn't necessarily translate into an improved description of the spectrum or gap compared to GF(0), as seen in Fig.~\ref{fig:compare_to_dynamic} and Tab.~\ref{tab:gw100}. This is because there is an imbalance in the implicit description of the ground state compared to the excitations, and it lacks a consistent truncation for effective error cancellation in energy differences (noting however that comparison to traditional CCSD is not an appropriate comparison given the difference in their ansatz for the excitations, but we do not think that this accounts for the majority of the discrepancy). For this reason, all moments are computed with the projections of ${\bar H}_N$ into the relevant space to be consistent with the EOM-CCSD description, and include the GF($0^\prime$) results only as an interesting comparison and approach to spectral properties from ground state information.

This approach is formally equivalent to the EKT-1 approach\cite{Ernzerhof2009, Welden2015, DiSabatino2021, Lee2021},
where the extended Koopman's theorem (EKT) is employed to produce ionisation potentials and electron affinities using the first two (zeroth and first) moments of the CC Green's function for the hole and particle sectors,
respectively.
Both approaches construct a spectrum consisting of the same number of poles
($N$ in each of the hole and particle sector,
where $N$ is the number of orbitals),
and both methods similarly conserve the first two moments, giving the same spectrum.
However, we note that the higher order EKT-$n$ are {\em not} equivalent to the higher moment constraints of GF($n$) which we detail in this work.

\subsection{GW100 test set} \label{sec:GW100}

\begin{figure}[htb!]
    \centering
    \subfloat[
        \centering
        Ionisation potential
    ] {
        \includegraphics[width=0.49\textwidth]{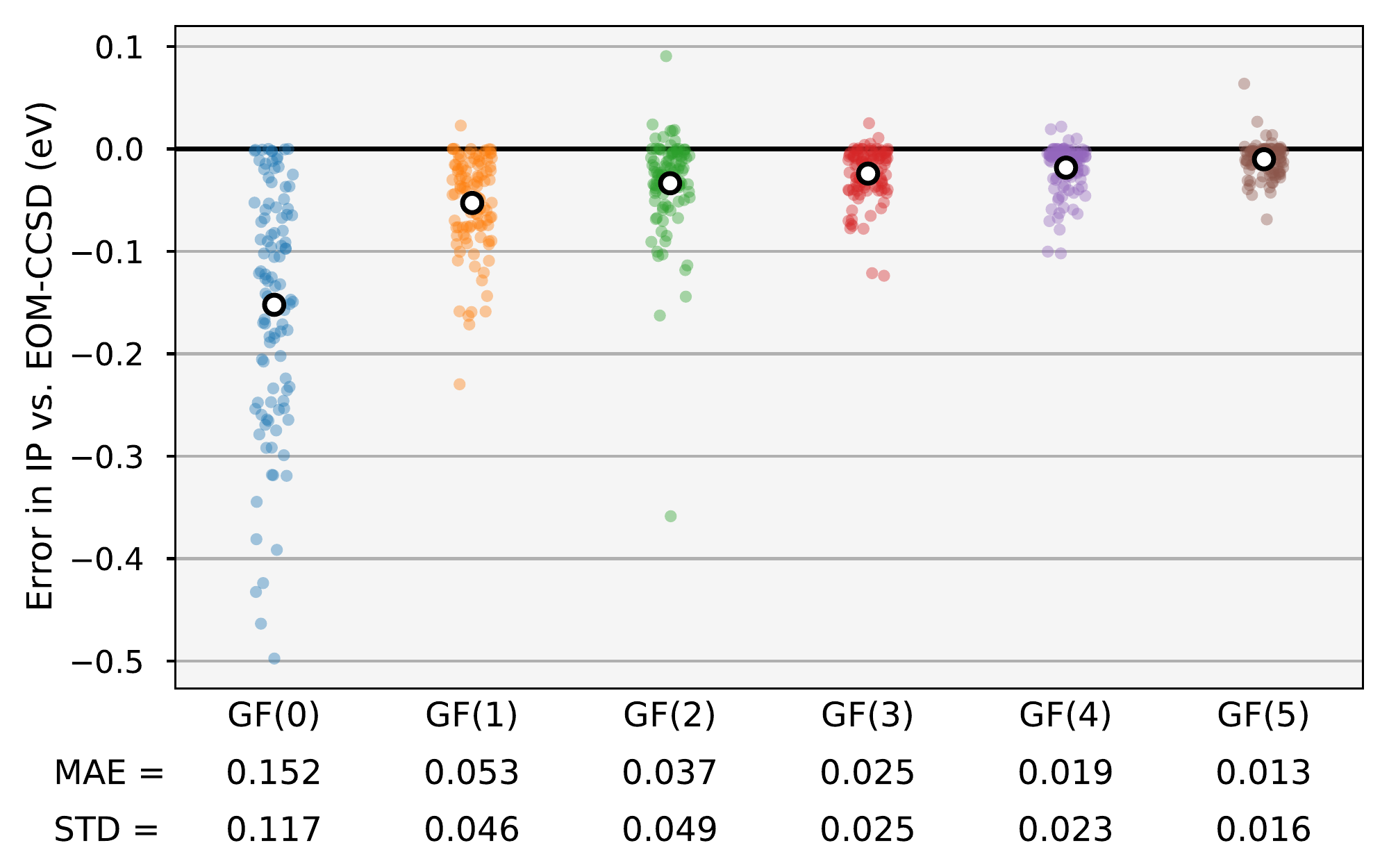}
    }
    \subfloat[
        \centering
        Electron affinity
    ] {
        \includegraphics[width=0.49\textwidth]{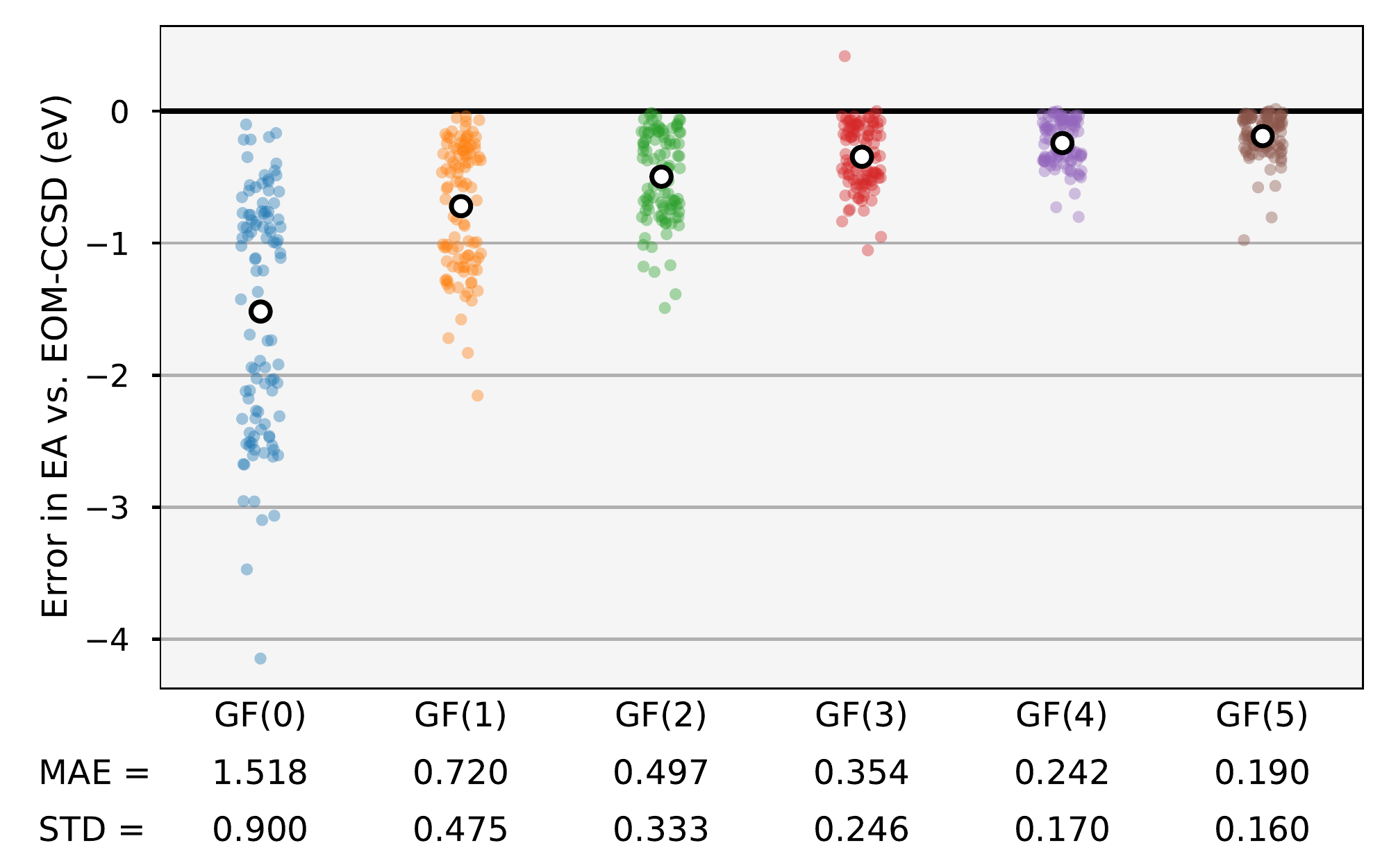}
    }
    \caption{
        Convergence of the moment-conservation approach to approximate the IP/EA-EOM-CCSD excitations for (a) ionisation potential and (b) electron affinity over the GW100 benchmark test set
        with a def2-TZVPP basis set. The white circle defines the mean absolute error at each order, with the standard deviations in these quantities also given below the plot (in units of eV).
        Excitations with a weight of $<0.1$ in the physical space were rejected in order to compare only quasiparticle-like excitations.
        The EOM-CCSD reference values were calculated using the \texttt{PySCF} package\cite{PySCF2017, PySCF2020}.
    }
    \label{fig:gw100_vs_ccsd}
\end{figure}
\begin{table}[htb!]
    \setlength{\tabcolsep}{10pt}
    \centering
    \begin{tabular}{l c c c c}
        \hline
        & \multicolumn{2}{c}{\textbf{IP}} & \multicolumn{2}{c}{\textbf{EA}} \\
        \textbf{Method} & \textbf{MAE} & \textbf{STD} & \textbf{MAE} & \textbf{STD} \\
        \hline
        HF & 0.627 & 0.708 & 0.982 & 0.470\\
        PBE & 3.996 & 1.226 & 3.071 & 0.899\\ \\
        GF(0$^\prime$)$^*$ & 0.226 & 0.197 & 1.461 & 0.904\\ \\
        GF(0) & 0.187 & 0.165 & 1.528 & 0.889\\
        GF(1) & 0.100 & 0.122 & 0.730 & 0.467\\
        GF(2) & 0.084 & 0.114 & 0.507 & 0.326\\
        GF(3) & 0.079 & 0.113 & 0.365 & 0.245\\
        GF(4) & 0.078 & 0.114 & 0.254 & 0.174\\
        GF(5) & 0.072 & 0.111 & 0.202 & 0.169\\ \\
        EOM-CCSD & 0.066 & 0.108 & 0.050 & 0.084\\
        ADC(2) & 0.661 & 0.949 & 0.331 & 0.474\\
        AGF2$^\dagger$ & 0.437 & 0.407 & 0.222 & 0.160\\
        $G_0W_0$@HF$^\ddagger$ & 0.307 & 0.263 & 0.188 & 0.206\\
        $G_0W_0$@PBE$^\ddagger$ & 0.665 & 0.328 & 0.243 & 0.248\\
        \hline
    \end{tabular}
    \caption{
        Mean absolute errors (MAE) and standard deviations (STD) of the GW100 test set for a series of methods,
        compared to references values at the $\Delta$CCSD(T) level of theory.
        Shown are the errors at mean-field Hartree--Fock (HF) and PBE density functional theory approximations, as well as the different CCSD GF moment truncations,
        along with the target EOM-CCSD errors and a number of additional methods employed in quantum chemistry to compute charged excitations. \newline
        $^*$The GF(0$^\prime$) data is missing the largest 10 systems due to memory constraints in storing the 2RDM (which could be alleviated with a more efficient implementation).
        $^\dagger$The AGF2 data is missing hexafluorobenzene and xenon due to convergence issues \cite{Backhouse2021}.
        $^\ddagger$The $G_0W_0$ data is missing the 8 systems which require an effective core potential (ECP) which was not available in the $\texttt{Fiesta}$ code used to compute them at the time of data collection \cite{Blase2011a, Blase2011b, Duchemin2020}.
        The $\Delta$CCSD(T) reference IP and EA values were calculated using the $\texttt{ORCA}$ code \cite{Neese2012, Neese2020}.
        All other values were computed using the PySCF programming package \cite{PySCF2017, PySCF2020}. 
    }
    \label{tab:gw100}
\end{table}
%
%
%
For broader and more reliable conclusions as to the convergence and accuracy of the GF moment expansion, Fig.~\ref{fig:gw100_vs_ccsd} shows the error in the ionisation potential (IP) and electron affinity (EA) of molecules in the GW100 test set compared to the state-specific convergence of the respective `infinite-moment' limit of IP/EA-EOM-CCSD values \cite{VanSetten2015}.
The systems were treated with a def2-TZVPP basis set with the corresponding effective core potential (ECP) applied to Rb, Ag, I, Cs, Au and Xe to treat core electrons.
The errors shown do not reflect the intrinsic error of the EOM-CCSD approximation for these excitations, and quantify only the convergence of the approximation to the lowest-energy excitations from the block Lanczos recursion of Sec.~\ref{sec:MomCCGF} to the true EOM-CCSD poles. Since the accuracy of this approximation relative to the inherent error of EOM-CCSD is important in judging the rate of convergence, in Table~\ref{tab:gw100} we show the aggregated errors between the methods and an accurate $\Delta$CCSD(T) benchmark, to include the intrinsic error compared to other perturbative and DFT quantum chemical methods for excitations.

We find errors in the first EA over this test set of Fig.~\ref{fig:gw100_vs_ccsd} to be significantly larger than those of the IP, by an order of a magnitude. This is rationalized by the fact that with any sufficiently large basis set, there will typically be many more virtual orbitals than occupied,
and therefore there are considerably more poles in the virtual part of the spectrum, which will also cover a substantially larger energy window.
As a result, the convergence of any \emph{single} excitation (e.g. the EA) is harder to achieve with the GF moments defined as integrated over all excitations, and with a larger number of true poles compressed into the same number as defining the occupied pole resolution for the same moment order. We note that there is no necessity in the algorithm to define the same moment truncation for the occupied and virtual spaces, and it is simple to specify a higher order in the IP or EA spaces as desired (including entirely excluding the IP or EA excitations if not of interest).

The intrinsic error (compared to $\Delta$CCSD(T)) of the EOM-CCSD method over the GW100 test set is 0.066 eV for the IP and 0.050 eV for the EA, which characterizes the aim for convergence of the moment expansion in these quantities.
Even with a single iteration of the recursion, defining GF(1),
the mean absolute error introduced by the moment approximation to the excitations is already less than this intrinsic error for the IP,
however the larger errors for the EA mean that the intrinsic error is not met up to GF(5).
Despite this,
for both the IP and EA there is a clear and systematic convergence in the error with increasing numbers of moments.
In comparison to other perturbative approximations to excitations in quantum chemistry,
even at GF(0) the error in IP compared to CCSD(T) outperforms ADC(2),
AGF2, and $G_0W_0$ due to its accurate approximation to the superior underlying EOM-CCSD description. For the GF(5) approximation, the error in the EA is similar to the AGF2 method, and only surpassed by the $G_0W_0$@HF (and EOM-CCSD) approximations, noting that this improvement of $G_0W_0$ does not extend to a PBE reference due to the strong reference dependence of the $G_0W_0$ approximation \cite{Bintrim2021}. 

In defining these specific frontier excitation energies rather than the overall spectrum, we note that the presented algorithm can suffer from the appearance of low-weighted excitations in the spectrum of Eq.~\ref{eq:compressed_spectrum} around the Fermi energy,
but which do not `adiabatically connect' to physical excitations in the exact limit of EOM-CCSD.
While these spectra will still conserve the input moments, the lack of imposing any explicit excitation structure does not preclude these erroneous excitations, which correspond to poles which are almost entirely located in the `external' space, and are only weakly coupled to the `physical' space.
Given their low weight and character far away from traditional frontier quasiparticles, these excitations can generally be identified and removed with an appropriate threshold when aiming to characterize individual frontier excitations. We therefore apply a threshold of a spectral weight of 0.1 in order to define the IP and EA here, which helps remove a small number of these erroneous excitations.

Another point to note is the potential for complex eigenvalues of $\mathbf{\bar{H}}_\mathrm{tri}$ with small imaginary components, manifesting in non-causality in the spectrum of Eq.~\ref{eq:compressed_spectrum}, which is not precluded in CC theory \cite{Thomas2021}. The potential for these appears to increase as one increases the number of conserved (non-Hermitian) CCSD moments.
In the case of the GF(2) IPs of Fig.~\ref{fig:gw100_vs_ccsd}, there exists a significant outlier (the water molecule),
whose spectrum correctly finds an excitation energy at an appropriate energy,
but where it does not correctly assign a good weight to this excitation and which is therefore removed by this threshold. These numerical aspects emphasise that care need to be taken when using the approach for the convergence of state-specific excitations in some cases, with the particular strength of the method in efficiently and systematically resolving the overall spectral intensities over large energy ranges.


\section{Conclusions}

We have presented an efficient and systematically improvable approximation for obtaining the coupled cluster Green's function which is rapidly convergent to the full resolution over all frequencies in a non-state-specific fashion, and requiring significantly fewer floating-point operations than more traditional approaches.
The solver is based on a modified block Lanczos recursion to directly conserve the particle and hole moments of a resulting Green's function, further adapted for use with the non-Hermitian spectral moments of CC theory.
We have shown an application of the approach for construction of full frequency Green's functions with spectral moments applied from level of CC singles and doubles, however stress that the recursion can equally well be applied to other levels of theory.
The convergence has been demonstrated using a series of full-frequency spectra compared to a traditional implementation,
and by demonstrating the convergence in the first ionisation potential and electron affinities of a large and diverse benchmark set.

\vtwo{The generality of the algorithm to build Green's functions via moment constraints will mean that future work will look to apply this approach using other methods to compute the GF moments,
including different levels of coupled cluster\cite{Bartlett2007}, stochastic methods \cite{PhysRevB.98.085118,PhysRevLett.118.176403} and $GW$ \cite{Bintrim2021}.}
Additionally,
the numerical stability and maintenance of biorthogonality within the recursion will be looked to be improved,
allowing effective convergence to the full-frequency limit and individual excitations.
The present algorithm should also be readily applicable to \textit{ab initio} solids using a $k$-space resolution,
where it can be combined with interpolation schemes to efficiently produce well-resolved spectra in solid state systems. Finally, adaptations towards optical excitations within a similar `static' moment-based framework will also be explored as alternative perspective for efficient spectral methods \cite{Scott2021}.


\ifjcp\else
    \newpage
\fi

\section*{Acknowledgements}

The authors sincerely thank Xavier Blase and Ivan Duchemin for help with obtaining $G_0W_0$ results over the GW100 test set, while G.H.B also thanks Emanuel Gull for illuminating discussions on Hamburger and Stieltjes moment problems.
G.H.B. also gratefully acknowledges support from the Royal Society via a University Research Fellowship as well as funding from the European Research Council (ERC) under the European Union’s Horizon 2020 Research and Innovation Programme (Grant Agreement
759063).
We are grateful to the UK Materials and Molecular Modelling Hub,
which is partially funded by the EPSRC (EP/P020194/1),
for computational resources.
Additionally, this work was funded under the embedded CSE programme of the ARCHER2 UK National Supercomputing Service (http://www.archer2.ac.uk).

\ifjcp
    \section*{References}
%
\else
    \bibliographystyle{achemso}
\providecommand{\latin}[1]{#1}
\makeatletter
\providecommand{\doi}
  {\begingroup\let\do\@makeother\dospecials
  \catcode`\{=1 \catcode`\}=2 \doi@aux}
\providecommand{\doi@aux}[1]{\endgroup\texttt{#1}}
\makeatother
\providecommand*\mcitethebibliography{\thebibliography}
\csname @ifundefined\endcsname{endmcitethebibliography}
  {\let\endmcitethebibliography\endthebibliography}{}

\fi

\section{Appendix}

\subsection{Derivation of first-order moments from RDMs} \label{app:moment_derivation}

For an arbitrary Hamiltonian with one- and two-body operators
\begin{align}
    \hat{H}
    &=
    \hat{h}
    +
    \hat{v}
    \\
    &=
    \sum_{pq}
    h_{pq}
    \des_{p}
    \cre_{q}
    +
    \frac{1}{2}
    \sum_{pqrs}
    v_{pqrs}
    \des_{p}
    \des_{q}
    \cre_{s}
    \cre_{r}
    .
\end{align}
One can derive an expression for the first-order hole and particle moments in terms of the one- and two-body reduced density matrices.
The first moment of the hole Green's function can be written (for a variational and Hermitian theory) as
\begin{align}
    T^{\mathrm{(h, 1)}}_{tu}
    &=
    \expval{\des_{t} \hat{H} \cre_{u}}
    \\
    &=
    \expval{\des_{t} [\hat{H}, \cre_{u}]}
    +
    \expval{\des_{t} \cre_{u}}
    E_0
    .
\end{align}
Using the identity
\begin{align}
    [\des_{p} \cre_{q}, \cre_{u}]
    &=
    -\cre_{q} \delta_{pu}
    ,
\end{align}
we can expand the one-body contribution to the commutator as
\begin{align}
    [\hat{h}, \cre_{u}]
    &=
    -\sum_{pq}
    h_{pq}
    \cre_{q}
    \delta_{pu}
    ,
\end{align}
and thus
\begin{align}
    \expval{\des_{t} [\hat{h}, \cre_{u}]}
    &=
    -\sum_{pq}
    h_{pq}
    \des_{t}
    \cre_{q}
    \delta_{pu}
    \\
    &=
    -\sum_{q}
    h_{uq}
    \des_{t}
    \cre_{q}
    \\
    &=
    -\sum_{q}
    h_{uq}
    \gamma_{tq}
    .
\end{align}
Using the identity
\begin{align}
    [\des_{p} \des_{q} \cre_{s} \cre_{r}, \cre_{u}]
    &=
    \des_{p} \cre_{s} \cre_{r} \delta_{qu}
    -
    \des_{q} \cre_{s} \cre_{r} \delta_{pu}
    ,
\end{align}
we can expand the two-body contribution to the commutator as
\begin{align}
    [\hat{v}, \cre_{u}]
    &=
    \frac{1}{2}
    \sum_{pqrs}
    v_{pqrs}
    \des_{p}
    \cre_{s}
    \cre_{r}
    \delta_{qu}
    -
    \frac{1}{2}
    \sum_{pqrs}
    v_{pqrs}
    \des_{q}
    \cre_{s}
    \cre_{r}
    \delta_{pu}
    ,
\end{align}
and thus
\begin{align}
    \expval{\des_{t} [\hat{v}, \cre_{u}]}
    &=
    \frac{1}{2}
    \sum_{pqrs}
    v_{pqrs}
    \des_{t}
    \des_{p}
    \cre_{s}
    \cre_{r}
    \delta_{qu}
    -
    \frac{1}{2}
    \sum_{pqrs}
    v_{pqrs}
    \des_{t}
    \des_{q}
    \cre_{s}
    \cre_{r}
    \delta_{pu}
    \\
    &=
    \frac{1}{2}
    \sum_{prs}
    v_{purs}
    \des_{t}
    \des_{p}
    \cre_{s}
    \cre_{r}
    -
    \frac{1}{2}
    \sum_{qrs}
    v_{uqrs}
    \des_{t}
    \des_{q}
    \cre_{s}
    \cre_{r}
    \\
    &=
    \frac{1}{2}
    \sum_{prs}
    \mel{pu}{}{rs}
    \Gamma_{tprs}
    .
\end{align}
Combining the one- and two-body terms for the shifted Hamiltonian $\hat{H}_N = \hat{H} - E_0$,
we arrive at an expression for the first-order hole moment
\begin{align}
    T^\mathrm{(h, 1)}_{tu}
    &=
    -\sum_{q}
    h_{uq}
    \gamma_{tq}
    +
    \frac{1}{2}
    \sum_{prs}
    \mel{pu}{}{rs}
    \Gamma_{tprs}
\end{align}
A similar derivation can be found for the $m=1$ moment of the particle Green's function, as
\begin{align}
    T^\mathrm{(p,1)}_{tu}
    &=
    \expval{
        \cre_{t}
        \hat{H}
        \des_{u}
    }
    \\
    &=
    \expval{
        \cre_{t}
        [\hat{H}, \des_{u}]
    }
    +
    \expval{\cre_{t} \des_{u}}
    E_0
    .
\end{align}
Using the identity
\begin{align}
    [
        \des_{p}
        \cre_{q}
        ,
        \des_{u}
    ]
    &=
    \des_{q}
    \delta_{pu}
    ,
\end{align}
we can expand the one-body contribution to the commutator as
\begin{align}
    [
        \hat{h}
        ,
        \des_{u}
    ]
    &=
    \sum_{pq}
    h_{pq}
    \des_{q}
    \delta_{pu}
    ,
\end{align}
and thus
\begin{align}
    \expval{
        \cre_{t}
        [
            \hat{h}
            ,
            \des_{u}
        ]
    }
    &=
    \sum_{pq}
    h_{pq}
    \cre_{t}
    \des_{q}
    \delta_{pu}
    \\
    &=
    \sum_{q}
    h_{uq}
    \cre_{t}
    \des_{q}
    \\
    &=
    \sum_{q}
    h_{uq}
    (
        \delta_{tq}
        -
        \des_{q}
        \cre_{t}
    )
    \\
    &=
    \sum_{q}
    h_{uq}
    (
        \delta_{tq}
        -
        \gamma_{tq}
    )
    .
\end{align}
Using the identity
\begin{align}
    [
        \des_{p}
        \des_{q}
        \cre_{s}
        \cre_{r}
        ,
        \des_{u}
    ]
    &=
    \des_{p}
    \des_{q}
    \cre_{s}
    \delta_{ru}
    -
    \des_{p}
    \des_{p}
    \cre_{r}
    \delta_{su}
    ,
\end{align}
we can expand the two-body contribution to the commutator as
\begin{align}
    \expval{
        \cre_{t}
        [
            \hat{v}
            ,
            \des_{u}
        ]
    }
    &=
    \frac{1}{2}
    \sum_{pqrs}
    v_{pqrs}
    \cre_{t}
    \des_{p}
    \des_{q}
    \cre_{s}
    \delta_{ru}
    -
    \frac{1}{2}
    \sum_{pqrs}
    v_{pqrs}
    \cre_{t}
    \des_{p}
    \des_{q}
    \cre_{r}
    \delta_{su}
    \\
    &=
    \frac{1}{2}
    \sum_{pqs}
    v_{pqus}
    \cre_{t}
    \des_{p}
    \des_{q}
    \cre_{s}
    -
    \frac{1}{2}
    \sum_{pqr}
    v_{pqru}
    \cre_{t}
    \des_{p}
    \des_{q}
    \cre_{r}
    .
\end{align}
Using the identity
\begin{align}
    \cre_{t}
    \des_{p}
    \des_{q}
    \cre_{s}
    &=
    (
        \delta_{pt}
        -
        \des_{p}
        \cre_{t}
    )
    \des_{q}
    \cre_{s}
    \\
    &=
    \delta_{pt}
    \des_{q}
    \cre_{s}
    -
    \des_{p}
    \cre_{t}
    \des_{q}
    \cre_{s}
    \\
    &=
    \delta_{pt}
    \des_{q}
    \cre_{s}
    -
    \des_{p}
    (
        \delta_{qt}
        -
        \des_{q}
        \cre_{t}
    )
    \cre_{s}
    \\
    &=
    \delta_{pt}
    \des_{q}
    \cre_{s}
    -
    \delta_{qt}
    \des_{p}
    \cre_{s}
    +
    \des_{p}
    \des_{q}
    \cre_{t}
    \cre_{s}
    ,
\end{align}
we can further expand this as
\begin{align}
    \nonumber
    \expval{
        \cre_{t}
        [
            \hat{v}
            ,
            \des_{u}
        ]
    }
    &=
    \frac{1}{2}
    \sum_{pqs}
    v_{pqus}
    \delta_{pt}
    \des_{q}
    \cre_{s}
    -
    \frac{1}{2}
    \sum_{pqs}
    v_{pqus}
    \delta_{qt}
    \des_{p}
    \cre_{s}
    \\ \nonumber
    &+
    \frac{1}{2}
    \sum_{pqs}
    v_{pqus}
    \des_{p}
    \des_{q}
    \cre_{t}
    \cre_{s}
    -
    \frac{1}{2}
    \sum_{pqr}
    v_{pqru}
    \delta_{pt}
    \des_{q}
    \cre_{r}
    \\
    &+
    \frac{1}{2}
    \sum_{pqr}
    v_{pqru}
    \delta_{qt}
    \des_{p}
    \cre_{r}
    -
    \frac{1}{2}
    \sum_{pqr}
    v_{pqru}
    \des_{p}
    \des_{q}
    \cre_{t}
    \cre_{r}
    \\ \nonumber
    &=
    \frac{1}{2}
    \sum_{qs}
    \mel{tq}{}{us}
    \gamma_{qs}
    -
    \frac{1}{2}
    \sum_{ps}
    \mel{pt}{}{us}
    \gamma_{ps}
    \\
    &+
    \frac{1}{2}
    \sum_{pqs}
    \mel{pq}{}{us}
    \Gamma_{pqst}
    .
\end{align}
The first two terms are identical within a sign via permutational symmetry of the two-electron integrals under exchange of the dummy labels $p$ and $q$,
leading to
\begin{align}
    \expval{
        \cre_{t}
        [
            \hat{v}
            ,
            \des_{u}
        ]
    }
    =
    \sum_{qs}
    \mel{tq}{}{us}
    \gamma_{qs}
    +
    \frac{1}{2}
    \sum_{pqs}
    \mel{pq}{}{us}
    \Gamma_{pqst}
    .
\end{align}
Combining the one- and two-body terms for a normal-ordered Hamiltonian we arrive at an expression for the first-order particle moment
\begin{align}
    \nonumber
    T^\mathrm{(p, 1)}_{tu}
    &=
    \sum_{q}
    h_{uq}
    (
        \delta_{tq}
        -
        \gamma_{tq}
    )
    +
    \sum_{qs}
    \mel{tq}{}{us}
    \gamma_{qs}
    \\
    &+
    \frac{1}{2}
    \sum_{pqs}
    \mel{pq}{}{us}
    \Gamma_{pqst}
    .
\end{align}
These expressions are in agreement with those of Ref.~\onlinecite{Lee2021}.

\end{document}